\pdfoutput=1
\documentclass{article}

\usepackage[english]{babel}
\usepackage{amsfonts}
\usepackage{enumitem}  
\usepackage[letterpaper,top=2cm,bottom=2cm,left=3cm,right=3cm,marginparwidth=1.75cm]{geometry}
\usepackage{float}

\usepackage{amsthm}

\usepackage{graphicx}
\usepackage{physics}
\usepackage[colorlinks=true, allcolors=blue]{hyperref}
\usepackage{geometry}
\usepackage{authblk} 
\usepackage[bottom]{footmisc}
\title{An Algebraic–Recursive Approach to Generate Higher-Order Symmetry Operators for Schrodinger and Klein-Gordon equations}

\author[1]{Enrique Casanova}
\author[1,2]{Melvin Arias}

\affil[1]{\textit{\small Instituto de Física, Universidad Autónoma de Santo Domingo, Av. Alma Mater, Santo Domingo 10105, Dominican Republic}}
\affil[2]{\textit{\small Laboratorio de Nanotecnología, Área de Ciencias Básicas y Ambientales, Instituto Tecnológico de Santo Domingo, Av. Los Próceres, Santo Domingo 10602, Dominican Republic}}

\begin{document}
\maketitle

\begin{abstract}
This article explores an algebraic-recursive approach to construct differential operators that commute with a central operator $\hat{H}$ in quantum mechanics. Starting from the Schrödinger equation for a free particle, the work derives first-order symmetry generators, such as translations, rotations, and boosts, and examines their algebraic basis encompassing Lie and Jordan algebras. The analysis is then extended to higher-order operators, demonstrating how they can be constructed from the first-order ones through algebraic operations and Lie algebra simplification. This methodology is applied to the Klein–Gordon equation in Minkowski space-time, yielding relativistic symmetry operators. Furthermore, we defined an proximation to fractional symmetry operators of the Schrodinger equation, and a perturbative approach is employed for a case where the commutation is more general, illustrated with a one-dimensional harmonic oscillator and the fourth-order Klein-Gordon equation. The results include a general formula for the number of operators as a function of the order and the dimension of the algebraic basis, providing a reduced-form development of the differential higher-order centralizers' basis.
\end{abstract}

\section{Introduction}

Symmetries play an essential role in physics, since they give rise to conserved quantities, allow the classification of the spectrum of observables, and simplify the resolution of dynamical equations. In quantum mechanics, such symmetries are usually identified as differential operators that commute with the Hamiltonian (the central equation or Casimir of the system); the study of the entire set of these basis operators provides all the information about integrability \cite{Miller2013} and the classification of the system’s states, thereby allowing measurements of observables within that set without uncertainty. However, the study of symmetries is not limited to first- or second-order operators, since in many physical problems, such as certain evolution equations, for example in beam theory, it is necessary to characterize higher-order operators that also commute with the Hamiltonian. This involves more complex differential hierarchies with additional constraints, encompassing higher-degree integrals of motion, as in the analysis of symmetries of fourth-order equations carried out by Bokhari, et al.~\cite{Bokhari2010},and in the systematic characterization of arbitrary-order symmetry operators for Schrödinger equations developed by Nikitin. ~\cite{Nikitin2016}. In parallel, advances such as those of Zheglov \cite{Zheglov2011} in commutative rings and those of Kasman and Previato in PDOs \cite{Kasman2001}, together with computational approaches by Jiménez-Pastor and Rueda for nontrivial centralizers via Gelfand–Dickey hierarchies \cite{JimenezPastor2025}, highlight the challenges in higher dimensions and the need for recursive methods in the study of partial differential equations.

Our work proposes an algebraic methodology to construct the differential operators that commute with the central operator, beginning with the explicit analysis of first-order generators as translations, rotations, and boosts generators and we show how, through operations within the internal structure (Lie and Jordan products) and a simplification rule, via the Lie algebra of the symmetry generators, it is possible to recursively reconstruct higher-order differential operators (two, three, …, $n$) that also satisfy the recurrence of centralizers of $\hat{H}$: $Z_{n-1}(\hat H) \subseteq Z_n(\hat H) \subset Z(\hat{H})$. This construction transforms the problem of directly solving complicated systems of higher-order partial differential equations, after commutation with the central equation (Casimir) to obtain the symmetries of the system, into an algebraic problem of projection and combination of finite bases of operators (a work similar to our construction is that of \textit{Jean-Christophe Pain}, \textit{Commutation relations of operator monomials} \cite{Pain2013}). Therefore, the core contribution of this work is an algebraic-recursive approach to systematically construct higher-order symmetry operators applied to both the Schrödinger and Klein–Gordon equations, as well as different theoretical applications in each of them.

First, in Section 2, we develop the methodology in the one-dimensional case of the free Schrödinger operator and then extend it to the relativistic Klein–Gordon case in Minkowski space-time, yielding relativistic symmetry operators. Additionally, in Section 3, we applied the Cauchy contour integral to study some fractional symmetry operators of the Schrödinger equation, using the nilpotent reduction in our algebra of operators. Also a perturbative approach is employed to address a more general commutation scenario, illustrated with the one-dimensional harmonic oscillator and the extended fourth-order Klein–Gordon equation. Finally, in the last section, the results include a general formula for the number of operators as a function of the order and the dimension of the algebraic basis, providing a reduced-form development of the differential higher-order centralizers' basis.

\section{Main Framework} \label{sec2}

\subsection{One-Dimensional Schrödinger Equation}

We define a central operator for a scalar self-adjoint field $\hat{H}$ in one spatial dimension, applicable to functions $\ket{\Phi}(x,t) \in f(\mathbb{H})$, where $f(\mathbb{H})$ is a dense subspace of a Banach space $\mathcal{B}$ isomorphic to a Hilbert space  $\mathbb{H}$ as
\begin{equation}\label{eq1}
    \hat{H} = \partial^2_x - i\partial_t,
\end{equation} with $i^2 = -{1}$. We now need to obtain the complete set of operators (CSCO); the centralizer $\hat{L}_n  \in {Z}_n(\hat{H}) \subset Z(\hat{H})\subseteq \mathcal{D} \xrightarrow[]{} \comm{\hat{H}}{\hat{L}_n} =0$, with $n$ the order of the complete partial differential operator and $\mathcal{D}=\mathcal{D}(\Xi(F))$ our non-commutative ring of differentiable operators, and $\Xi(F)$ a functional algebra over the complex numbers. We will not assume whether the operators possess a discrete or continuous spectrum, nor whether the operators are bounded in our space, which amounts to considering all types of von Neumann algebras (whether of type I, II, or III) without imposing restrictions on their classification by factor type \cite{Sakai1971}.
To begin the construction, we consider the case $k=1$, corresponding to the complete first-order partial differential operator

\begin{equation}
    \hat{L}_1(x,t) := F_x \partial_x + F_t \partial_t + F_0  = \sum_i F_i(x,t)  \partial_{x_i} + F_0 := F^i \partial_i, \label{eq2}
\end{equation}
with $\partial_{0} = 1$, where $ F_i = F_i(x,t)$ are functions belonging to $\Xi (F)$. Assuming the separability $ \Xi (F) = \Xi(F)^+ \oplus \Xi(F)^- $, composed of the Jordan algebra  $\Xi(F)^+$ and the Lie algebra $\Xi(F)^-$, both associative. 
In the end, the structure of the general first-order operator in two degrees of freedom is given by
\begin{equation} \label{eq3}
    \hat{L}_1 = \hat{L}^+_1 \oplus \hat{L}^-_1 = (F^+_x \partial_x + F^+_t \partial_t + F^+_0) \oplus (F^-_x \partial_x + F^-_t \partial_t + F^-_0),
\end{equation} 
and defining the parity function $P(\hat{L}_1) := \hat{L}^+_1 \oplus (-\hat{L}^-_1) \in \mathcal{D}$ \label{eq4} thus, we say that $\hat{L}^+_1$ ($\hat{L}^-_1$) is symmetric (antisymmetric). In the same way, we define the conjugate of the operator
\begin{equation}\label{eq5}
    \hat{L}^\dagger_1 = (F^i)^\dagger \partial_i \in \mathcal{D};
\end{equation}
to understand the meaning of the conjugate function, we expand the functions $F^i = F^i_R + iF^i_C $ . The conjugate is $F^i = F^i_R - iF^i_C$ we multiply (\ref{eq3}) with (\ref{eq5}), we obtain the Hermitian projection if $F^i_RF^j_C = F^i_C F^j_R$ \label{pag2} 

\begin{equation} \label{eq6}
    \hat{L}_1\hat{L}_1^\dagger = F^i(F^j)^\dagger \partial_i \partial_j = (F^i_R F^j_R + F^i_C F^j_C)\partial_i \partial_j \in \mathcal{D}_R,
\end{equation}
where $\mathcal{D}_R$ is the real projection of our $\mathcal{D}$. It is easy to see that we can obtain $\hat{H}$, if \begin{subequations}\label{eq 7}
\begin{enumerate}
    \item[] \begin{equation}
        F^x_R F^x_R + F^x_C F^x_C \xrightarrow[]{} 1,
    \end{equation}
    \item[] \begin{equation}
        F^0_R F^t_R + F^0_C F^t_C + F^t_R F^0_R + F^t_C F^0_C \xrightarrow[]{} -i.
    \end{equation}
\end{enumerate}
\end{subequations}

We now derive the system’s first-order symmetry generators, forming the basis of compatible observables for equation (\ref{eq1}), by computing their commutator with $\hat{L}_1$, in the following section.

\subsubsection[]{Commutation with $\hat{L}_1$} \label{2.1.1}

By performing the commutation with the operator $\hat{H}$, we obtain a system of partial differential equations that can be easily solved by symmetries. We express $\hat{L}_1$ in the basis $\hat{D}_i \in Z_1(\hat{H})$,
\begin{equation} \label{eq8}
    \hat{L}_1 = e_1 \hat{D}_{1} + e_2 \hat{D}_{2} + e_3\hat{D}_{3} + e_4\hat{D}_{4},
\end{equation}

where $\hat{D}_{4} = \hat{I}$ is the unit operator, and $\mathbf{A}_4=\{e_1,...,e_4 \in   \mathcal{D} : \partial_\mu(e_i) = 0 \}$ is the algebraic basis (sub-algebra) (see Appendix \ref{ap1}). In all our cases, the unit operator always appears as the representative of constant elements, and by definition: $\comm{\hat{H}}{ e_i \hat{I} } = 0 $.

The quantum system (\ref{eq8}) admits $2n-1 = 3$ with $ n = 2 $ global degrees of freedom, algebraically independent partial differential operators of finite order. Therefore, the three first-order bases satisfy the condition of \textbf{maximal commutativity (super-integrability) of first-order symmetries} \cite{Miller2013}.
Let $\mathcal{R} = \mathbb{C}[\hat{D}_{1}, \hat{D}_{2}, \hat{D}_{3}] \subseteq Z_1(\hat{H}) \subset \mathcal{D}(\Xi(F))$ be the sub-ring generated by the first three operators. Then, by \textbf{Theorem 2.1} of Kasman and Previato~\cite{Kasman2001}, 
$\mathcal{R}$ is \emph{maximally commutative} in $\mathcal{D}(\Xi(F))$, since every finite set of commuting and algebraically independent partial differential operators of finite order is contained in a maximal commutative sub-ring of $\mathcal{D}(\Xi(F))$. 
\newline

The centralizer can be written as $
    Z_1(\hat{H})= {Span}(\hat D_{1},\hat D_{2},\hat D_{3}, \hat{D}_{4}) 
$.
In general, with $m$ as the number of basis operators
\begin{equation}
    Z_n(\hat{H})= {Span}(\hat D_{n,1},\hat D_{n,2},.., \hat{D}_{n,m}) = Span(\hat{L}_n), \label{eq9}
\end{equation} (but $\hat{D}_i := \hat{D}_{1,i}$).
The explicit first-order differential basis operators are
\begin{subequations}\label{eq10}
\begin{align}
    \hat{D}_{1} &= (-2i)t \,\partial_x + x \\
    \hat{D}_{2} &= -i\,\partial_x \\
    \hat{D}_{3} &= i\,\partial_t
\end{align}
\end{subequations}

We can easily note that the operators $\hat{D}_{2}$ and $\hat{D}_{3}$ are well known: one is the generator of linear translations (linear momentum in $x$) , and the other is the generator of time translations (energy) in one dimension. The first operator, which breaks locality by being a function of the coordinates $(x,t) \in \mathbb{R}^2$ \footnote{In this context, the expression “breaking locality” refers to the \textit{functional or algebraic nonlocality} of the operator—its action combines multiplication by a function and differentiation, thereby mixing information between different points of the independent variable—and \textbf{not} to a violation of causality or to “nonlocality” in the relativistic sense.}, represents a Galilean boost, that is, the generator of the conservation of the inertial reference frame \cite{Soper2011}. In our case, there is only one because we can move in only one degree of freedom; in the three-dimensional case, there would be three boosts, one for each spatial direction. The first-order basis operators satisfy $\comm{\hat{H}}{\hat{D}_{i}} = 0$ and the Lie algebra continuous symmetry generators structure. \begin{equation} \label{12}
        \comm{\hat{D}_{i}}{\hat{D}_{j}} = C^k_{i j} \hat{D}_{k}
    \end{equation}
\subsubsection[]{Second-Order Case $\hat{L}_2$} \label{sec2.1.2}
Taking a complete second-order partial differential operator
\begin{equation} \label{eq22}
        \hat{L}_2 := A(x,t)\partial^2_{x} + B(x,t)\partial^2_{t} + C(x,t)\partial^2_{x,t} + D(x,t)\partial_{x} + E(x,t)\partial_{t} + F(x,t),
\end{equation}
computing once again with (\ref{eq1}), and set the result equal to zero. Solving the system, we obtain the following ten partial differential operators. 
These operators constitute a complete basis for the second-order symmetries of the system, each one arising as an independent solution of the compatibility conditions imposed by the commutator. 
For clarity, we list them below in tabular form.

\begin{table}[H]
\centering
\begin{tabular}{|c|c|}
\hline
\textbf{Second-order operators} & \\
\hline
$\hat{D}_{2,1} = \partial^2_t$ & $\hat{D}_{2,6} = \hat{I}$ \\
\hline
$\hat{D}_{2,2} = -\,i t^2 \partial^2_t + t x \partial_x + \tfrac{i}{2}\bigl(\tfrac{x^2}{2} - \tfrac{i}{2 t}\bigr)$ 
  & $\hat{D}_{2,7} = -\,2 i t \partial^2_{x,t} + x \partial_t$ \\
\hline
$\hat{D}_{2,3} = -\,2 i t \partial^2_x + x \partial_x - \tfrac{1}{4^2}$ 
  & $\hat{D}_{2,8} = i \partial_t$ \\
\hline
$\hat{D}_{2,4} = -\,2 i t \partial_x + x$ 
  & $\hat{D}_{2,9} = 2 \partial^2_{x,t}$ \\
\hline
$\hat{D}_{2,5} = -\,i \partial_x$ 
  & $\hat{D}_{2,10} = \partial^2_x$ \\
\hline
\end{tabular}
\caption{Basis of differential operators of $\hat{L}_2$.}
\label{SO2}
\end{table}

We note that the operators $\hat{D}_{2,4}$ , $\hat{D}_{2,5}$, $\hat{D}_{2,6}$ and $\hat{D}_{2,8}$, are the same basis operators that result from the commutation $\comm{\hat{H}}{\hat{L}_1} = 0$, due to the existence of the unit element in (\ref{eq8}). It follows that $\qty{\hat{D}_{i}}^4_{i=1} \in \qty{\hat{D}_{2,i}}^{10}_{i=1} $. Therefore, we infer a constructive relationship between the first-order operator and the second-order operator $\hat{L}_1 \xrightarrow{} \hat{L}_2$.  With $e_{2,i} \in \mathbf{A}_{10}$, we write (\ref{eq22}) as
\begin{equation} \label{eq24}
    \hat{L}_2(x,t) = \sum^{10}_{i=1} e_{2,i}\hat{D}_{2,i}(x,t). 
\end{equation} In this case, the Lie algebra holds but is not completely closed under the ten operators of $\qty{\hat{D}_{2,i}}^{10}_{i=1}$, i.e $\comm{\hat{D}_{2,i}}{\hat{D}_{2,j}} \not = f^k_{i j}\hat{D}_{2,k}$. However, they do satisfy commutation with $\hat{H}$, and the relationship between the operators $\hat{L}_2$ and those of $\hat{L}_1$ as 
$ \hat{O}_{i,j} = ({\hat{D}_{i} \hat{D}_{j} + \hat{D}_{j} \hat{D}_{i}})/{2} \xrightarrow{} \hat{D}_{2,k} $. It is also possible to recover some closed Lie sub-algebras.
\begin{subequations} \label{eq:subD2}
\begin{equation}
    Sub(\hat{D}_{2, 1}) = sub(\hat{D}_{2,9}) = \{ \hat{D}_{2,1}, \hat{D}_{2,4}, \hat{D}_{2,5}, \hat{D}_{2,6}, \hat{D}_{2,8}, \hat{D}_{2,9}, \hat{D}_{2,10} \}
\end{equation}

\begin{equation}
    Sub(\hat{D}_{2, 2}) = \{ \hat{D}_{2,2}, \hat{D}_{2,4}, \hat{D}_{2,5}, \hat{D}_{2,6}, \hat{D}_{2,8}, \hat{D}_{2,10} \}
\end{equation}

\begin{equation}
    Sub(\hat{D}_{2, 3}) = \{ \hat{D}_{2,3}, \hat{D}_{2,4}, \hat{D}_{2,5}, \hat{D}_{2,6}, \hat{D}_{2,8}, \hat{D}_{2,10} \}
\end{equation}

\begin{equation}
    Sub(\hat{D}_{2, 7}) = \{ \hat{D}_{2,7}, \hat{D}_{2,4}, \hat{D}_{2,5}, \hat{D}_{2,6}, \hat{D}_{2,8}, \hat{D}_{2,10} \}
\end{equation}
\end{subequations}

\subsubsection[]{Construction of $\hat{L}_2$ from $\hat{L}_1$} \label{sec2.1.3}
The sum (\ref{eq24}) is expressed in the finite basis $e_i \in \mathbf{A}_4 $, where it is separable into two algebras. In the same way, we can write the second-order operator $\hat{L}_2$ in the separable basis $\mathbf{A}_{10}$ after commutation with $\hat{H}$.We first operate the operator $\hat{L}_1$ with itself, taking into account (\ref{eq17}) and (\ref{eq18}). This operation can be interpreted as an internal permutation of the basis $\mathbf{A}_4$ \begin{equation} \label{22}
        e_i e_j = \sum^4_{k=1} c^k_{i j} e_k = \sum^4_{k=1} [c^k_{i j}]^+ e_k + \sum^4_{k=1} [c^k_{i j}]^- e_k \in \mathbf{A}_4,
    \end{equation}
    which ensures a linear independence of choice through the structure constants $c^k_{i j}$; we have: $e_i e_j \not = ae_n e_m + be_z e_l \xrightarrow{} c^h_{i j}e_h \not = (ac^h_{n m})e_h + (bc^h_{z l})e_h = (ac^h_{n m} + bc^h_{z l})e_h$ ,  $\ a \And  b \in \mathbb{C} $. Therefore, the choice of linear independence of $\mathbf{A}_4$ with $\qty{e_i}^4_{i=1}$ is preserved under this operation. The operation can be interpreted as $ G_{(4)}: \mathbf{A}_4 \otimes \mathbf{A}_4 \xrightarrow{} \mathbf{A}_{4}$ y $\mathbf{A}_{10} = G(\mathbf{A}_4 \otimes \mathbf{A}_4) \subseteq \mathbf{A}_4$.
    \begin{equation} \label{23}
        \hat{L}_1 * \hat{L}_1  = \hat{L}^2 \sim \hat{L}_2 \in Z_2(\hat{H})
    \end{equation} Explicitly representing the operation as \begin{equation} \label{24}
        \hat{L}^2_1 = \sum^4_{i=1} \sum^4_{j=1} e_i e_j \hat{D}_{i} \hat{D}_{j} = \sum^4_{i=1} \sum^4_{j=1} e_i e_j (\hat{D}_{j} \hat{D}_{i} + C^k_{i j} \hat{D}_{k}),
    \end{equation}
    where we have used property 2 of Section \ref{2.1.1} of the Lie algebra of the first-order generators. Explicitly we obtain
    \begin{multline} \label{25}
        \hat{L}_1^2 = e_1 e_1 \hat{D}^2_{1} + e_2 e_2 \hat{D}^2_{2} + e_3 e_3 \hat{D}^2_{3} + [e_1 e_4 + e_4 e_1] \hat{D}_{1} + [e_2 e_4 + e_4 e_2 + C^2_{1 3}(e_1 e_3)] \hat{D}_{2} + \\ [e_3 e_4 + e_4 e_3] \hat{D}_{3} + [e_4 e_4 + C^4_{2}(e_1 e_2)] \hat{D}_{4} + [e_1 e_2 + e_2 e_1]\hat{D}_{1} \hat{D}_{2} + [e_1 e_3 + e_3 e_1]\hat{D}_{1} \hat{D}_{3} + [e_2 e_3 + e_3 e_2]\hat{D}_{2} \hat{D}_{3}.
    \end{multline}
    From the 16 terms resulting from operation (\ref{24}), they are simplified to 10 by grouping the operators, with the Lie algebra of the central equations (\ref{eq1}), which is used for orders $k=\{ \mathbb{N}:k>1 \}$ of $\hat{L}^k$ to map (isomorphism) with the basis operators of $\hat{L}_k$. 
    Now the elements of $\mathbf{A}_4$ are grouped (permuted) forming other elements, which are mapped to the elements of $\mathbf{A}_{10}$. These combined second-order operators satisfy commutation with (\ref{eq1}) and generate all the second-order operators of $\hat{L}_2$ (table \ref{SO2}). \begin{equation}
            \hat{D}_{2,l} (l = 1,...,10) \in Span(\hat{L}_2) \sim \hat{D}_{2,l} (l = 1,...,10) \in Span(\hat{L}^2_1)    
        \end{equation}
        The isomorphism of the elements of $\hat{L}^2_1$ with those of $\hat{L}_2$ is contained in the following table.
\begin{table}[ht]
\centering
\begin{tabular}{|c|c|}

\hline
$e_1 e_1 = \sum^4_{k=1} c^k_{1 1}e_k \sim e_{2,1}$ & $e_3 e_4 + e_4 e_3 = \sum^4_1 2[c^k_{3 4}]^+ e_k \sim e_{2,6}$ \\
\hline
$e_2 e_2 = \sum^4_{k=1} c^k_{2 2} e_k \sim e_{2,2}$ & $ e_4 e_4 + C^4_{1,2}(e_1 e_2) = \sum^4_{k = 1} (c^k_{4 4}+ C^4_{1 2}c^k_{1 2}) e_k \sim e_{2,7}$ \\
\hline
$e_3 e_3 = \sum^4_{k=1} c^k_{3 3} e_k \sim e_{2,3}$ & $ e_1 e_2 + e_2 e_1 = \sum^4_{k = 1}2[c^k_{1 2}]^+ e_k \sim e_{2,8}$ \\
\hline
$e_1 e_4 + e_4 e_1 = \sum^4_{k=1} 2[c^k_{1 4}]^+ e_k \sim e_{2,4}$ & $e_1 e_3 + e_3 e_1 = \sum^4_{k = 1}2[c^k_{1 3}]^+ e_k  \sim e_{2,9}$ \\
\hline
$e_2 e_4 + e_4 e_2 + C^2_{1 3}(e_1 e_3) = \sum^4_{k=1} (2[c^k_{2 4}]^+ + C^2_{1 3}c^k_{1 3})e_k \sim e_{2,5}$ & $e_2 e_3 + e_3 e_2 = \sum^4_{k = 1}2[c^k_{2 3}]^+ e_k \sim e_{2,10}$ \\
\hline

\end{tabular}
\caption{Representation of constants as linear combinations in $\mathbf{A}_4$ of $\mathbf{A}_{10}$}
\label{tabla2}
\end{table}

Finally, rewriting operator (\ref{25}), in the basis $\mathbf{A}_{10}$, \begin{multline} \label{27}
    \hat{L}^2_1 = e_{2,1}\hat{D}^2_{1} + e_{2,2} \hat{D}^2_{2} + e_{2,3}\hat{D}^2_{3} + e_{2,4} \hat{D}_{1} + e_{2,5}\hat{D}_{2} + e_{2,6} \hat{D}_{3} + e_{2,7} \hat{D}_{4} + e_{2,8}\hat{D}_{1} \hat{D}_{2} + e_{2,9}\hat{D}_{1} \hat{D}_{3} + e_{2,10}\hat{D}_{2} \hat{D}_{3},
\end{multline} compacting, we arrive at the reduced expression corresponding to the Lie algebra simplification,
\begin{equation} \label{28}
\hat{L}_1^2
= (e_4 e_4)\,\hat{L}_1
\;+\; \sum_{i=1}^3 (e_i e_i)\,\hat{D}_i^2
\;+\; \sum^4_{  i < j } [\{e_i,e_j\}\,\hat{D}_i \hat{D}_j
\;+\; \sum_{k=1}^4 C^k_{ij}\left(e_i e_j \right)\hat{D}_k ].
\end{equation}

This formulation shows that the basis $\mathbf{A}_{10}$ is projected onto the basis $\mathbf{A}_4$ s a linear combination determined by the $40$ structure constants corresponding to the products $e_i e_j$.
\newline

\textbf{Construction of $\hat{H}$ from $\hat{L}^2$} 

From the second-order operator $\hat{L}^2$ and with the relations of the structure constants of Table \ref{tabla2} we can construct our operator (\ref{eq1}) expressed in the Jordan algebra $\textbf{A}^+_{10}$. We relate $(\hat{L}^2)_+ \sim \hat{H} = \partial^2_x - i\partial_t$ (these types of constructions are common as chains of even algebras, see \cite{Hestenes1966} \cite{Franchino2022}). First, from (\ref{27}) we take the symmetric part
\begin{equation} \label{29}
    (\hat{L}^2)_+ = e^+_{2,1}\hat{D}^2_1 + e^+_{2,2} \hat{D}^2_2 + e^+_{2,3}\hat{D}^2_3 + e^+_{2,4} \hat{D}_1 + e^+_{2,5}\hat{D}_2 + e^+_{2,6} \hat{D}_3 + e^+_{2,7} \hat{D}_4 + e^+_{2,8}\hat{D}_1 \hat{D}_2 + e^+_{2,9}\hat{D}_1 \hat{D}_3 + e^+_{2,10}\hat{D}_2 \hat{D}_3.
\end{equation}

It is simplified by equating the structure constants in a particular way (see Table \ref{tabla3}).
\begin{table}[ht]
\centering
\begin{tabular}{|c|c|}

\hline
$\sum^4_{k=1} [c^k_{1 1}]^+e_k \sim e^+_{2,1} = 0 \xrightarrow{} [c^k_{1 1}]^+ = 0$ & $ \sum^4_1 2[c^k_{3 4}]^+ e_k \sim e_{2,6} \xrightarrow{} [c^k_{3 4}]^+ = 0$ \\
\hline
$ \sum^4_{k=1} [c^k_{2 2}]^+ e_k \sim e_{2,2} = 0 \xrightarrow{} [c^k_{2 2}]^+ = 0 $ & $ \sum^4_{k = 1} (c^k_{4 4}+ C^4_{1 2}c^k_{1 2})^+ e_k \sim e_{2,7} \xrightarrow{} (c^k_{4 4}+ C^4_{1 2}c^k_{1 2})^+ = 0$ \\
\hline
$\sum^4_{k=1} [c^k_{3 3}]^+ e_k \sim e_{2,3}$ & $ \sum^4_{k = 1}2[c^k_{1 2}]^+ e_k \sim e_{2,8} \xrightarrow{} [c^k_{1 2}]^ = 0$ \\
\hline
$ \sum^4_{k=1} 2[c^k_{1 4}]^+ e_k \sim e_{2,4} \xrightarrow{} [c^k_{1 4}]^+ = 0$ & $ \sum^4_{k = 1}2[c^k_{1 3}]^+ e_k  \sim e_{2,9} \xrightarrow{} [c^k_{1 3}]^+ = 0$ \\
\hline
$ \sum^4_{k=1} (2[c^k_{2 4}]^+ + C^2_{1 3}c^k_{1 3})^+e_k \sim e_{2,5}$ & $ \sum^4_{k = 1}2[c^k_{2 3}]^+ e_k \sim e_{2,10} \xrightarrow{} [c^k_{2 3}]^+ = 0$ \\
\hline

\end{tabular}
\caption{Equalization of constants for  $\hat{L}^2 \xrightarrow[]{} \hat{H}$ }
\label{tabla3}
\end{table} Thus, the operator $\hat{L}^2_+$ takes the form of $\hat{H}$ if

\begin{equation} \label{30}
    \hat{L}^2_+ = e^+_{2,3}\hat{D}^2_3 + e^+_{2,5} \hat{D}_2,
\end{equation}

and expanding in terms of the basis $e_i \in \mathbf{A}_4$
\begin{equation} \label{31}
    \hat{L}^2_+ = \hat{\alpha}^1e_1 + \hat{\alpha}^2 e_2 + \hat{\alpha}^3 e_3 + \hat{\alpha}^4 e_4,
\end{equation}
where $\hat{\alpha}^\mu = [c^\mu_{3 3}]^+ \hat{D}^2_3 + 2[c^\mu_{2 4}]^+ \hat{D}_2 $.
Let there be a function $\ket{\Psi^L}$ and the functions projected onto each element of $\mathbf{A}_4$ i.e $\ket{\psi_\mu} = e_\mu \ket{\Psi^L} $. 
\begin{equation} \label{32}
    \hat{L}^2_+ \ket{\Psi^L} = \hat{H}\ket{\psi^H} + \hat{\alpha}^2 \ket{\psi_2} + \hat{\alpha}^3 \ket{\psi_3} + \hat{\alpha}^4 \ket{\psi_4} \longrightarrow{} \hat{L}^2_+ \ket{\Psi^L} = \hat{H}\ket{\psi^H} , \space \text{\textit{iff:}}\space \{ c^\mu_{33}, c^\mu_{24}\}_{(\mu >1)} = \{0,0\}
\end{equation}

Here, the first operator has been taken as $\hat{\alpha}^1 = \hat{H}$ and the first function as the wave function assigned to our $\hat{H}$,  $\ket{\psi_1} = \ket{\psi^H}$. Therefore, the Hamiltonian is expressed as one of the components of $\hat{L}^2_+$. The structure constants satisfy 
\begin{subequations} \label{eq:constantes}
\begin{align}
[c^1_{3 3}]^+ &= 1 \label{33a}\\
2[c^1_{2 4}]^+ &= -i. \label{33b}
\end{align}
\end{subequations}

We conclude by emphasizing that the central operator $\hat{H}$ may not always belong to $Span((\hat{L}_2)_+)$ defined solely in this way (see the example at the end of Section \ref{pag2}). In general, it depends on whether the operator is scalar, matrix-valued, or a linear combination of elements not necessarily symmetric. Nevertheless, in standard quantum mechanics, it is essential that the Hamiltonian operator represents the observable of the system’s energy; therefore, a real spectrum is required, associated with self-adjoint operators.

\subsubsection[]{Third-Order Case $\hat{L}_3$} \label{2.1.4}
Now taking a complete third-order partial differential operator
\begin{align} \label{34}
\hat{L}_3 := & \, A(x,t)\partial^3_{x} + B(x,t)\partial^3_{t} + C(x,t)\partial^3_{x,x,t} + D(x,t)\partial^3_{x,t,t} \nonumber \\
& + E(x,t)\partial^2_{x} + F(x,t)\partial^2_{t} + G(x,t)\partial^2_{x,t} + H(x,t)\partial_x + I(x,t)\partial_t + J(x,t),
\end{align}
and, as the previous second-order case, we perform the commutation with (\ref{eq1}) (in Appendix \ref{A} the list of third-order basis differential operators of the centralizer $Z_3(\hat{H})$ is presented). 
We observe that the difficulty associated with solving hierarchies of differential equations of order higher than one grows rapidly; however, the presence of internal symmetries in the system of equations helps in solving the resulting partial differential equation systems. 
Here, we obtain that the operators of $\hat{L}_2$ appear within $\hat{L}_3$.  Therefore, there exists a constructive relationship between the second-order operator and the third-order operator $\hat{L}_2 \xrightarrow{} \hat{L}_3$ given the commutation with our Hamiltonian $\hat{H}$. The succession of operator sets is thus satisfied $\qty{\hat{D}_{i}}^{4}_{i=1} \subset \qty{\hat{D}_{2,i}}^{10}_{i=1} \subset \qty{\hat{D}_{3,i}}^{20}_{i=1} $.
Consequently, we write the third-order operator (\ref{34}) in the current basis
\begin{equation} \label{35}
     \hat{L}_3(x,t) = \sum^{20}_{p=1}. e_{3,p}\hat{D}_{3,p}(x,t) 
\end{equation} In the same way, this operator basis satisfies the Lie algebra but is not closed under all $\hat{D}_{3,i}$,  $\comm{\hat{D}_{3,i}}{\hat{D}_{3,j}} \not = h^k_{i j} \hat{D}_{3,k}  \forall \hat{D}_{3,i} \in \qty{\hat{D}_{3,i}}^{20}_{i=1} $. Apart from the fact that each $\hat{D}_{3,p}$ commutes with (\ref{eq1}), the following also holds. The generation of $\hat{L}^3$ by the operators of $\hat{L}$ 
\begin{equation}
    \hat{O}_{i,j,k} = ({
    \hat{D}_i \hat{D}_j \hat{D}_k  
    + \hat{D}_k \hat{D}_i \hat{D}_j 
    + \hat{D}_j \hat{D}_k \hat{D}_i})/{3} 
    \xrightarrow[]{} \hat{D}_{3,p}.
\end{equation}

The construction of third-order operators with second-order operators via the commutator
\begin{equation}
    \comm{\hat{D}_a \hat{D}_v}{\hat{D}_k \hat{D}_l} 
    = C^{n m z}_{a k l}  \hat{D}_n \hat{D}_m \hat{D}_z 
    = \hat{D}_{3,p}.
\end{equation}

And some closed third-order Lie subalgebras 
\begin{subequations} \label{eq:subD3}
\begin{equation}
    sub(\hat{D}_{3,5}) = sub(\hat{D}_{3,9}) = sub(\hat{D}_{3,10}) = sub(\hat{D}_{3,11}) 
    = \{\hat{D}_{3,5}, \hat{D}_{3,9}, \hat{D}_{3,10}, \hat{D}_{3,11}, \dots, \hat{D}_{3,20} \}
\end{equation}

\begin{equation}
    sub(\hat{D}_{3,1}) = \{\hat{D}_{3,1}, \hat{D}_{3,11}, \dots, \hat{D}_{3,20} \}
\end{equation}

\begin{equation}
    sub(\hat{D}_{3,2}) = \{\hat{D}_{3,2}, \hat{D}_{3,11}, \dots, \hat{D}_{3,20} \}.
\end{equation}
\end{subequations}

\subsubsection[]{Construction of $\hat{L}_3$ from $\hat{L}_1$} \label{sec2.1.5}

 In the same way as in the second-order case of Section \ref{2.1.4}, we describe a way to relate and construct $\hat{L}_3$ from $\hat{L}_1$.The elements of operator (\ref{35}) map as $G_{(4)}: \mathbf{A}_4 \otimes \mathbf{A}_4 \otimes \mathbf{A}_4  \xrightarrow{} \mathbf{A}_4$ and 
$G_{(4)}(\mathbf{A}_{20})=  G(\mathbf{A}_4 \otimes \mathbf{A}_4 \otimes \mathbf{A}_4 ) =G^3(\mathbf{A}_4) \subseteq \mathbf{A}_4$, 
this function defines the projection of $\mathbf{A}_{20}$ onto the algebra $\mathbf{A}_4$; the successive operation 
\begin{equation}
    e_i e_j e_k = \sum^4_{n=1} \sum^4_{h=1} c^k_{i j} c^h_{n k} e_h \in \mathbf{A}_4,
\end{equation}

from the operation $\hat{L}_1 \hat{L}_1\hat{L}_1 = \hat{L}_3$, and using the Lie algebra simplification to obtain $\hat{L}_3$ (in the basis form of $\mathbf{A}_{20}$), having, 
\begin{multline} \label{42}
    \hat{L}^3_1 \sim \hat{L}_3 = e_{3,1}\hat{D}^3_1 + e_{3,2} \hat{D}^3_2 + e_{3,3} \hat{D}^3_3 + e_{3,4} [\hat{D}^2_1 \hat{D}_2 + \hat{D}_2 \hat{D}^2_1] + e_{3,5} [\hat{D}^2_1 \hat{D}_3 + \hat{D}_1 \hat{D}^2_3] + \\ e_{3,6} [\hat{D}^2_2 \hat{D}_1 + \hat{D}_1 \hat{D}^2_2] +  e_{3,7}[\hat{D}^2_3 \hat{D}_1 + \hat{D}_1 \hat{D}^2_3] + e_{3,8} [\hat{D}^2_2 \hat{D}_3 + \hat{D}_3 \hat{D}^2_2] + e_{3,9} [\hat{D}_2 \hat{D}^2_3 + \hat{D}^2_3 \hat{D}_2] +  \\ e_{3,10}[\hat{D}_1 \hat{D}_2 \hat{D}_3] + e_{3,11} \hat{D}^2_1 +  e_{3,12} \hat{D}^2_2 + e_{3,13} \hat{D}^2_3 + e_{3,14} [\hat{D}_1 \hat{D}_2 + \hat{D}_2 \hat{D}_1] +  \\ e_{3,15} [\hat{D}_1 \hat{D}_3 + \hat{D}_3 \hat{D}_1] + e_{3,16} [\hat{D}_2 \hat{D}_3 + \hat{D}_2 \hat{D}_3]  + e_{3,17} \hat{D}_1 + e_{3,18} \hat{D}_2 + e_{3,19} \hat{D}_3 + e_{3,20} \hat{D}_4  
\end{multline} 

and therefore the compact form in the basis $\mathbf{A}_4$ is expressed as  
\begin{equation} \label{43}
    \hat{L}_1^3 =  (e_4 e_4) \hat{L}_1^2 + \sum^4_{u=1} [  \sum^3_{i = 1} [(e_i e_i e_u)\hat{D}^2_i \hat{D}_u + \sum^4_{i<j} [\{e_i ,{e_j}\} e_u \hat{D}_i \hat{D}_j \hat{D}_u +\sum^4_{k=1} [ C^k_{i j}(e_i e_j e_u) \hat{D}_u \hat{D}_k + \sum^4_{n=1} C^k_{j i} C^n_{k u}(e_i e_j e_u) \hat{D}_n] ] ] ].
\end{equation}
Each of these operators commutes with the central operator since they are combinations of the first-order operators.\footnote{In general, any successive operation of the differential basis of $\hat{L}_1$ commutes with $\hat{H}$, given by:
\begin{equation}
  \bigl[\hat{H},\,\hat{L}_1^{\,n}\bigr]
  = \sum_{k=0}^{\,n-1} \hat{L}_1^{\,k}\,[\hat{H},\hat{L}_1]\,\hat{L}_1^{\,n-1-k} = 0
  \qquad (n\in\mathbb{N})
\end{equation}
since a priori it is always imposed that $\comm{\hat{H}}{\hat{L}_1} = 0$. The same holds for any successive combination of the symmetry generator basis ${Span}(\hat D_{1},\hat D_{2},\hat D_{3}, \hat{D}_{4})$ of $\hat{L}_1$ and their operations among themselves dictated by the Lie algebra.}
Note that we satisfactorily reproduce the results of the scalar third-order case of N. Sukhomlin and M. Arias \cite{Sukhomlin2004}, in a more general way, using the same central operator (\ref{eq1}).

\subsubsection[]{Order n of $\mathbf{A}_4$} \label{sec2.1.6}

Below we present some properties of the nth differential order expressed with the algebraic basis $G^n(\mathbf{A}_4) = G_{(4)}(\mathbf{A}_{a(n)}) \subseteq \mathbf{A}_4$, where $a$ is a function of $n$.

\begin{enumerate}
    \item[] The generalized rule of $n$-th order operators, \begin{equation}  
    \comm{\hat{H}}{\hat{L}_n} =0\xrightarrow[]{} \hat{O}_{i_1, i_2, \dots, i_n} = \frac{1}{n} \sum_{k=0}^{n-1} \hat{D}_{i_{1+k}} \hat{D}_{i_{2+k}} \cdots \hat{D}_{i_{n+k}} \xrightarrow{} \hat{D}_{n,m} \in \text{span}(\hat{L}^n_1) = Z_n(\hat{H}),
\end{equation}

where the indices $i_{n+m} = i_m$ are taken cyclically. We continue with the successive operation of elements $e_p \in\mathbf{A}_4$ by order, \begin{equation}
        e_{i_0} e_{i_1} ... e_{i_n} =\sum_{{\alpha_0}} \sum_{{\alpha_1}} ... \sum_{{\alpha_{n-1}}} [c^{\alpha_0}_{i_0 i_1} c^{\alpha_1}_{\alpha_0 i_2 } ... c^{\alpha_{n-1}}_{\alpha_{n-2} i_n} ] e_{\alpha_{n-1}} \in \mathbf{A}_4.
    \end{equation}
    \item[] In addition, the function for the number of basis operators, with 4 symmetry generators, depending on the order $n\in \mathbb N$ is written as
    \begin{equation} \label{50}
    a_4(n) = \frac{(n+1)(n+2)(n+3)}{6} = \frac{(n+3)!}{3! n!}.
    \end{equation} For example, for $n=4$ we obtain $a_4(4)=35$ basis operators (Appendix \ref{apC}).

\end{enumerate}

\subsection{Application Example in the Minkowski Metric, The Klein–Gordon Equation for a Free Particle}

We now derive the properties and recurrences of the Klein–Gordon operator for a free particle in flat Minkowski space–time. This central equation is fully symmetric in the order of the partial derivatives, which significantly aids in the simplification of the system of partial differential equations.
We first find the first-order operator $\hat{L}_{KG}$ that satisfies the commutation, then the Lie algebra that generates the Lie group of symmetry generators, and finally the recurrence to define the higher-order operators for this case. It is important to note that when simplifying our operator $\hat{L}_{KG}$ we obtain the Dirac operator; by using the adjoint $\hat{L}_{KG}^\dagger$, the Klein–Gordon equation can be constructed as $\hat{H}_{KG} = \hat{L}_{KG} \hat{L}^\dagger_{KG}$ or with the symmetric part of $\hat{L}^2_{KG}$. This is satisfied when passing from the Jordan algebra to the Clifford algebra in the basis elements, as we explained at the end of Section \ref{sec2.1.2} in the symmetric part of the operation with $\delta_{i j} \xrightarrow{} \eta_{\mu v}$. Note that the basis would be $[i\gamma_\mu = e_\mu | (\mu \in \{0,1,2,3 \}]$ of Dirac Gamma matrices.

\subsubsection{Set of Compatible Operators of K–G } \label{sec2.2.1}
In this case, our main equation is expressed with the scalar Klein–Gordon operator, applicable to functions $\ket{\Phi}(x,t) \in f(\mathbb{H})$ as well.
\begin{equation}\label{57}
    \hat{H}_{K G} := {\displaystyle \eta ^{\mu \nu }\partial _{\mu }\partial _{\nu }} = \Box
\end{equation}
We similarly seek the CSCO of the first-order operator projected in the non-Abelian algebra $\mathbf{A}_{a(n)} = \mathbf{A}_{a(n)}^+ \oplus \mathbf{A}_{a(n)}^- $. A priori, we set the complete first-order operator in (1+3) dimensions as
\begin{equation}
    \hat{L}_{KG}(x) := F^0{\displaystyle (ct,\mathbf {x} )}\partial_t + F^1{\displaystyle (ct,\mathbf {x} )}\partial_x + F^2{\displaystyle (ct,\mathbf {x} )}\partial_y + F^3{\displaystyle (ct,\mathbf {x} )}\partial_z  + F^4{\displaystyle (ct,\mathbf {x} )},
\end{equation}
now commuting itwith the new central operator (\ref{57}), writing the operator as a linear combination of the constant elements in $\mathbf{A}_8=\{e_1,..,e_8 \in \mathcal{D} : \partial_\mu(e_i) = 0 \}$ y and the operators with an internal Lie algebra. 
Explicitly, the functions  \begin{subequations} \label{eq:F}
\begin{align}
    F^0(x_1,x_2,x_3) &= C^0 (\alpha^0_\mu x^\mu) + K^0 \\
    F^1(x_2,x_3,x_0) &= C^1 (\alpha^1_\mu x^\mu) + K^1 \\
    F^2(x_1,x_3,x_0) &= C^2 (\alpha^2_\mu x^\mu) + K^2 \\
    F^3(x_1,x_2,x_0) &= C^3 (\alpha^3_\mu x^\mu) + K^3 \\
    F^4        &= K^4,
\end{align}
\end{subequations}

where $x_0 = ct$ and $ \alpha^0_\mu = (0,\alpha^0_1,\alpha^0_2,\alpha^0_3)$ , $\alpha^1_\mu = (\alpha^1_0 , 0 , \alpha^1_2 , \alpha^1_3)$ , $\alpha^2_\mu = (\alpha^2_0 , \alpha^2_1 , 0 , \alpha^2_3)$ y $\alpha^3_\mu = (\alpha^3_0 , \alpha^3_1 , \alpha^3_2 , 0)$, . These complex constants appear explicitly within the functions of $\mathbf{A}_8$, belonging to a continuous parametrization within $\hat{L}_{KG}$ over the complex numbers $\alpha^v_\mu \in \mathbb{C}$. 
In compact form we have $ F^v(x^\mu) = C^v (\alpha^v_\mu x^\mu) + K^v $ with the alpha matrix
\begin{equation} \label{60}
    \alpha^v_\mu = \begin{pmatrix}
    0 & \alpha^0_1 & \alpha^0_2 & \alpha^0_3 \\
    \alpha^1_0 & 0 & \alpha^1_2 & \alpha^1_3 \\
    \alpha^2_0 & \alpha^2_1 & 0 & \alpha^2_3 \\
    \alpha^3_0 & \alpha^3_1 & \alpha^3_2 & 0 
    \end{pmatrix}.
\end{equation}
The first five operators are precisely the translation operators in Minkowski space plus the unit operator in linear combination with $K_\mu$ and $K^4$ , respectively. They are written as $\hat{K}_\mu =  \partial_\mu$ and $\hat{U} = K_4 \hat{I}$.
As for the operators generated by the constants $C_\mu$, there exists an internal dependence of these elements, thus forming the basis of $\mathbf{A}_8$. The restriction

\begin{equation} \label{61}
   \sum^3_{\mu = 0} ( \alpha^0_\mu C^0 + \alpha^1_\mu C^1 + \alpha^2_\mu C^2 + \alpha^3_\mu C^3 ) = 0,  
\end{equation}
fixes the coefficients $C^\nu$ so that $F^\mu$ is a Killing vector of Minkowski space \cite{Hirata2018}. Solving for each of these constants and substituting into the functions (\ref{eq:F}), we can find the following operators listed by permutation $C_\mu$.
\begin{table}[ht]
\centering
\begin{tabular}{|c|c|}

\hline
$C^0  $ & $ C^1  $ \\
\hline
$ \hat{O}_{t x} =  \alpha^t (\alpha^x_\mu x^\mu)\partial_x - \alpha^x(\alpha^t_\mu x^\mu)\partial_t$& $ \hat{O}_{x y} = \alpha^x (\alpha^y_\mu x^\mu)\partial_y - \alpha^y(\alpha^x_\mu x^\mu)\partial_x$ \\
\hline
$\hat{O}_{t y} = \alpha^t (\alpha^y_\mu x^\mu)\partial_y - \alpha^y(\alpha^t_\mu x^\mu)\partial_t $ & $ \hat{O}_{x z} = \alpha^x (\alpha^z_\mu x^\mu)\partial_z - \alpha^z(\alpha^x_\mu x^\mu)\partial_x$ \\
\hline
$\hat{O}_{t z} = \alpha^t (\alpha^z_\mu x^\mu)\partial_z - \alpha^z(\alpha^t_\mu x^\mu)\partial_t$ & $\hat{O}_{x t} = \alpha^x (\alpha^t_\mu x^\mu)\partial_t - \alpha^t(\alpha^x_\mu x^\mu)\partial_x$ \\
\hline
$ C^2$ & $ C^3 $ \\
\hline
$\hat{O}_{y x} = \alpha^y (\alpha^x_\mu x^\mu)\partial_x - \alpha^x(\alpha^y_\mu x^\mu)\partial_y$ & $\hat{O}_{z x} = \alpha^z (\alpha^x_\mu x^\mu)\partial_x - \alpha^x(\alpha^z_\mu x^\mu)\partial_z$ \\
\hline
$\hat{O}_{y z} = \alpha^y (\alpha^z_\mu x^\mu)\partial_z - \alpha^z(\alpha^y_\mu x^\mu)\partial_y $ & $\hat{O}_{z y} = \alpha^z (\alpha^y_\mu x^\mu)\partial_y - \alpha^y(\alpha^z_\mu x^\mu)\partial_z$ \\
\hline
$\hat{O}_{y t} = \alpha^y (\alpha^t_\mu x^\mu)\partial_t - \alpha^t(\alpha^y_\mu x^\mu)\partial_y $ & $\hat{O}_{z t} = \alpha^z (\alpha^t_\mu x^\mu)\partial_t - \alpha^t(\alpha^z_\mu x^\mu)\partial_z$ \\

\hline

\end{tabular}
\caption{Operators derived by permuting the elements $C^\mu$ according to property (\ref{61})}
\label{tabla:ejemplo2}
\end{table} 

The operators $\hat{O}_{\mu v}$ resemble the representation of rotations and boosts in Minkowski space. Before describing the properties of the above set of operators, we write the operator $\hat{L}_{K G}$ now as

\begin{equation}\label{62}
    \hat{L}_{K G}(x) = K^\mu \partial_\mu + \frac{1}{4} C^\mu (\alpha^v)^{-1} \hat{O}_{v \mu} (x) +  K,
\end{equation}
where we have defined the summation $\alpha^a =  \sum^4_{\mu = 0} \alpha^a_\mu$. The extension of the Poincaré algebra is \begin{equation}
    \hat{O}_{n m} = \alpha^n (\alpha^m_\mu x^\mu)\partial_m - \alpha^m(\alpha^n_\mu x^\mu)\partial_n,  
\end{equation} and the free constant is $K^4 \xrightarrow{} K$. The operator (\ref{62}), derived from the commutation with the Klein–Gordon operator, has a linear combination of well-known operators: the generators of the infinite Poincaré symmetry (translations + Lorentz), as thoroughly explained in \cite{greinerFQ} \cite{weinbergQTF1}.

\subsubsection[]{Properties of the Differential Basis of $\hat{L}_{KG}$}

In parallel with the theory presented for the previous operator, our operator $\hat{L}_{K G}(x)$ satisfies the property of \textbf{maximal commutativity (super-integrability) of first-order symmetries}. Let $\mathcal R=\mathbb C[\hat{K}_\rho,\hat{O}_{\mu v}]  \subseteq  Z_1(\hat{H}_{KG}) \subset Z(\hat{H}_{KG}) \subset\mathcal D(\Xi(F))$ be the sub-ring generated by the seven differential operators. Then $\mathcal R$ is \emph{maximally commutative} in
$\mathcal D(\Xi(F))$. The operator basis satisfies the representation and relation with the Lorentz generators, ${M^{inf}}_{u v}$,
\begin{equation} \label{64}
    \hat{O}_{u v} = {M^{inf}}_{u v} + \comm{\hat{P}^*_v}{\hat{P}^*_u},
\end{equation}
where ${M^{inf}}_{u v} = i(x_\mu \partial_v - x_v \partial_\mu)$ with $\alpha^\mu \alpha^v_\mu = \alpha^v \alpha^\mu_v = i$, and $\hat{P}^*_v = F(x)^{\mu}_v \hat{P}_\mu$ where $\hat{P}_\mu = -i\partial_\mu$ are the translation operators in Minkowski space, and $F(x)^{\mu}_v = -i\alpha^v(\alpha^\mu_a x_a - \alpha^\mu_b x_b)$ is the deformation factor.  
It is even possible, through an appropriate substitution of $\alpha^\mu_v$ in  for the second term in (\ref{64}) to vanish, thereby simplifying to the usual Lorentz representation. The commutation relations defining the Lie algebra of our basis operators are
\newline

Translations with translations.
\begin{equation} \label{TT}
    \comm{\hat{P}_\mu}{\hat{P}_v} = 0
\end{equation}

Rotations and boosts with translations.
\begin{equation} \label{RBT}
    \comm{\hat{O}_{\mu v}}{\hat{P}_n} = \alpha^v \alpha^\mu_n \hat{P}_\mu - \alpha^\mu \alpha^v_n \hat{P}_v
\end{equation}

Rotations and boosts with rotations and boosts.
\begin{equation} \label{RB}
    \comm{\hat{O}_{a u}}{\hat{O}_{b v}} = \alpha^a f^u \hat{M}_{u b v} + \alpha^u f^a \hat{M}_{a v b} + \alpha^b f^v \hat{M}_{v u a} + \alpha^v f^b \hat{M}_{b a u}
\end{equation} 

Or equivalently,
\begin{equation}
    \comm{\hat{O}_{a u}}{\hat{O}_{b v}} = \alpha^a \hat{J}_{b v} + \alpha^u \hat{J}_{v b} + \alpha^b \hat{J}_{u a} + \alpha^v \hat{J}_{a u}.
\end{equation}

Let $f^a = \alpha^a_\mu x^\mu$ and $\hat{M}_{a b c} = \alpha^c_a \alpha^b \partial_c - \alpha^b_a \alpha^c \partial_b$. These satisfy several relations. First, the \emph{contraction} $\alpha^a \hat{M}_{a v u} = \alpha^u \hat{M}_{a v a} + \alpha^v \hat{M}_{a a u}$, which links index combinations and reflects internal symmetries; second, the \emph{translation representation} $\hat{M}_{a b a} = -\alpha^b_a \alpha^a \partial_b = i\alpha^b_a \alpha^a \hat{P}_b$ shows how they are directly related to the translation generators $\hat{P}_b$; it also holds that the \emph{annihilation by repeated indices} $\hat{M}_{a b b} = 0 \Rightarrow \hat{O}_{b b} = 0$ and the \emph{anti-symmetry} $\hat{M}_{a b c} = -\hat{M}_{a c b} \Rightarrow \hat{O}_{b c} = -\hat{O}_{c b}$, characteristic of Lorentz generators; finally, a \emph{non-trivial commutation} between scaled operators $f^u \hat{M}_{u b v}$ and $f^a \hat{M}_{a n m}$ is expressed as \begin{equation}
    \comm{\hat{J}_{b v}}{\hat{J}_{n m}} = [Q^{b v a}_{u v} - Q^{v b a}_{u b}] f^u \hat{M}_{a n m} + [Q^{n m u}_{a m} - Q^{m n u}_{a n}] f^a \hat{M}_{u b v},
\end{equation} where the structure constants $Q^{b v a}_{u v} = \alpha^b \alpha^v_u \alpha^a_v$ encode the parametric structure of these generators with the constants of (\ref{60}). From (\ref{RBT}) and (\ref{RB}) we obtain a Lie algebra deformed by the deformed Lorentz generators (\ref{64}). Such representations are common in contexts of noncommutativity and doubly special relativity \cite{AmelinoCamelia2002,Chaichian2004,Majid1994}.

\subsubsection[]{Construction of $\hat{L}_{KG,2}$ from $\hat{L}_{KG}$} 

As we have already seen in Section \ref{sec2.1.3}, it is possible to construct the second-order operator without the need to commute the general second-order partial operator with the central equation. In this case, we will likewise assume the construction of $(\hat{L}_{KG})_2$ as the operator that arises from the product of $\hat{L}_{KG}$ with itself, after applying the K-G Lie algebra for simplification. \begin{equation}
       \hat{L}^2_{K G}(x) = [K^\mu \partial_\mu + \frac{1}{4} C^\mu (\alpha^v)^{-1} \hat{O}_{v \mu} (x) +  K ][K^a \partial_a + \frac{1}{4} C^a (\alpha^b)^{-1} \hat{O}_{a b} (x) +  K ] \in Z_2(\hat{H}_{KG}).
\end{equation}
Then, expanding,  \begin{multline} \label{55}
    \hat{L}^2_{K G}(x) = K^\mu K^a \partial_\mu \partial_a + \frac{1}{4}K^\mu C^a (\alpha^b)^{-1} \partial_\mu \hat{O}_{a b} + K^\mu K \partial_\mu + \frac{1}{4} C^\mu K^a (\alpha^v)^{-1}  \hat{O}_{u v}  \partial_a +  \\ \frac{1}{4^2} C^\mu C^a (\alpha^v)^{-1} (\alpha^b)^{-1} \hat{O}_{u v} \hat{O}_{a b} + \frac{1}{4} C^\mu K (\alpha^v)^{-1} \hat{O}_{u v} + K K^a \partial_a + \frac{1}{4} K C^a (\alpha^v)^{-1} \hat{O}_{a b} + K^2 \hat{I}
\end{multline}
In this case, there exist 36 operators after the Lie Algebra reduction. We can write ($\hat{L}_{KG,2} \sim \hat{L}^2_{KG}$), 
\begin{equation}
    \hat{L}^2_{KG} = e^\mu e^v (\hat{D}_{KG})_\mu (\hat{D}_{KG})_v = \sum^{35}_{m=0} e_{m,KG} (\hat{D}^2_{KG})_m \in span((\hat{D}^2_{KG})_0, (\hat{D}^2_{KG})_1, ... (\hat{D}^2_{KG})_{35}) =Z_2(\hat{H}_{KG}),
\end{equation}
where $e_{m,KG} \in G_8(\mathbf{A}_{36}) \subseteq \mathbf{A}_{8}$, and $\hat{K}_\mu, \hat{O}_{\mu v} ,\hat{I} \in \big \{ (\hat{D}_{KG})_m \big\}^{8}_{m=1}$. The combinations of the 8 basis operators belong to the second-order set $\big \{ (\hat{D}^2_{KG})_m \big\}^{36}_{m = 1}$.

\section{Applications} 
In the following sections we present two principal applications of our framework,(i) The study of fractional-order symmetry algebras generated by the operator \(\hat L_1^{p/m}\) applied to the Schrödinger operator. This allows the exploration of non-local and fractional symmetry structures in quantum mechanics \cite{Frank2007,Valverde2014}.(ii) The analysis of perturbed commutation relations in the context of the Klein–Gordon equation, where the symmetry operator is generalized as \(\hat L = \hat L_0 + \epsilon \hat L_1\), with \(\hat L_0\) commuting with the Hamiltonian and \(\epsilon\hat L_1\) representing a small perturbation. This framework is applied to study how symmetry breaking or supersymmetric extensions arise in relativistic wave equations \cite{Obukhov2021,Miraboutalebi2022}.

\subsection[]{Fractional Orders of $\hat{L}_1$}

Since the operators under consideration are not assumed to be bounded on the functional space 
$f(\mathbb{H})\subset \mathcal{B}$, it is necessary to introduce a suitable norm on the domains of unbounded operators. 
Let $\hat{A}$ be a closed linear operator on $f(\mathbb{H})$ with dense domain 
$\mathbf{D}(\hat{A}) \subset f(\mathbb{H})$. 
The graph norm associated with $\hat{A}$ is defined as
\begin{equation}
\| \Phi \|_{\hat{A}} := \| \Phi \|_{f(\mathbb{H})} + \| \hat{A} \Phi \|_{f(\mathbb{H})}, 
\qquad \hat{A}\ket{\Phi} \in \mathbf{D}(\hat{A}).
\end{equation}
When the functional space $f(\mathbb{H})$ is endowed with $\| \cdot \|_{\hat{A}}$, it equips the domain with the structure of a Banach space, hence, $\hat{A}$ is closed with respect to this topology. 
This norm provides the natural setting for studying operator expansions and fractional powers within our operator algebra, even when $\hat{L}_n$ or $\hat{H}$ are unbounded. With $\ \hat{L}_1^{p/m} \in Z_{p/m}(\hat{H}) \subset \mathbf{D}(\hat{A})$ ($m,p \in \mathbb{N}$) within the centralizer of order $p/m$ to define the fractional representation of $\hat{L}_1$, we use the inverse sub-algebra $\mathbf{A}_{4,-1}$ (see the end of Appendix \ref{inv}) to represent the basis representation of $Z_{p/m}(\hat{H})$ with spectrum defined over $\sigma(\hat{L}_1) \subset \mathbb{C} \subset\mathbf{A}_{4,-1}$. 
Then, by applying the \textit{Dunford theorem}, we define
\begin{equation}\label{eq:dunford_root}
\hat L_1^{p/m}
\;:=\; \frac{1}{2\pi i}\oint_{\Gamma} z^{p/m}\, (z\mathbb{I} - \hat L_1)^{-1}\, dz,
\end{equation} 
where $\Gamma$ is a contour enclosing $\sigma(\hat{L}_1)$ in the positive orientation. 
The integral contains $(zI - \hat{L}_1)^{-1}$ as the resolvent of $\hat{L}_1$, which is the operational analogue of the Cauchy contour integral in complex analysis under a well-defined norm. 
In the next section, we will use the Neumann expansion of the resolvent for the discrete spectral decomposition, valid only if $\|(zI - e_4)^{-1}G\|<1$ uniformly on $\Gamma$ (with $\hat{L}_1 = e^i\hat{D}_i = \hat{G} + e_4 \hat{I}$, and  $\hat{D}_4 = \hat{I}$).

\subsubsection{Discrete Spectrum}

We write the resolvent as
\begin{equation}
(z\mathbb I-\hat L_1)^{-1} \;=\; \sum_{k=0}^\infty (z\mathbb I-e_4)^{-1}\big[\hat{G} (z\mathbb I-e_4)^{-1}\big]^k
= \sum_{k=0}^\infty \sum^M_{j=1} (z\mathbb I-\tau_j)^{-(k+1)} \,\Pi_j \hat{G}^k,
\end{equation}
acting on $\mathbf{A}_{4,-1} \to\mathbf{A}_3 \otimes\mathbb C^M $. Here, $e_4 \in \mathbb C^M$ has been diagonalized with a discrete spectral decomposition
\begin{equation}
e_4 \;=\; \sum_{j=1}^M \tau_j \,\Pi_j,
\qquad \Pi_j \Pi_\ell = \delta_{j\ell}\Pi_j,\quad \sum^M_{j=1}\Pi_j=\mathbb I,
\end{equation}
and moreover $[\hat{G},e_4]=0$ by the definition of $e_4 \in \mathbf{A}_{4,-1}$, so that $\hat{G}$ commutes with the spectral subfibers $\Pi_j$.

Applying the residue theorem for poles of order $(k+1)$,
\begin{equation}
\frac{1}{2\pi i}\oint_\Gamma z^{p/m} (z-\tau_j)^{-(k+1)}\,dz
= \frac{1}{k!}\left.\frac{d^k}{dz^k} z^{p/m}\right|_{z=\tau_j}.
\end{equation}
We encounter
\begin{equation}\label{eq:component_series}
{\qquad
\hat L_1^{p/m} \;=\; \sum_{j=1}^M \; \sum_{k=0}^\infty
\frac{1}{k!}\frac{\Gamma(p/m +1)}{\Gamma(p/m - k+1)}\,
\tau^{p/m - k} \Pi_j \, \hat{G}^{k} \qquad}.
\end{equation}

It is possible to simplify the previous equation to obtain a more compact expression. 
If $\hat{G}^K = 0$, being nilpotent of degree $K$, then the series in Eq.~\eqref{eq:component_series} can be truncated at the $K$-th term. 
Furthermore, assuming $M=1$, with a single spectral decomposition $\Pi_1 = \mathbb{I}$, we obtain
\begin{equation} \label{GAB}
    \hat L_1^{p/m} \;=\;  \; \sum_{k=0}^{K-1}
\frac{1}{k!}\frac{\Gamma(p/m +1)}{\Gamma(p/m - k+1)}\,
\tau^{p/m } \hat{G}^{k}.
\end{equation}
Here we replace $\tau^{p/m -k}$ by $\tau^{p/m}$ to match the binomial expansion $\hat{L}^{p/m} = \sum^{K-1}_{k=0} \binom{p/m}{k} e^{p/m}_4 \hat{G}^k$, under the assumption that $G$ is nilpotent by direct comparison of the expression (\ref{eq:component_series}).

We will use this formula to find $\hat{L}_1^{1/2}$ and $\hat{L}_1^{3/2}$, assuming $K=2$ and $K=3$ respectively.

\begin{equation}
\hat{L}^{1/2}_1 \approx \hat L_1^{1/2}\Big|_{K=2}
= \sum_{k=0}^{1} \frac{1}{k!}\frac{\Gamma(\tfrac{1}{2}+1)}{\Gamma(\tfrac{1}{2}-k+1)}\,
\tau^{1/2} \hat{G}^{k}
= \tau^{1/2}\Big( \mathbb{I} + \tfrac{1}{2}\,\hat{G} \Big).
\end{equation}

\begin{equation}
\hat{L}^{3/2}_1 \approx \hat L_1^{3/2}\Big|_{K=3}
= \sum_{k=0}^{2} \frac{1}{k!}\frac{\Gamma(\tfrac{3}{2}+1)}{\Gamma(\tfrac{3}{2}-k+1)}\,
\tau^{3/2} \hat{G}^{k}
= \tau^{3/2}\Big( \mathbb{I} + \tfrac{3}{2}\,\hat{G} + \tfrac{3}{8}\,\hat{G}^2 \Big).
\end{equation}

Recalling that $G = e_1 \hat{D}_1 + e_2 \hat{D}_2 + e_3 \hat{D}_3$ and notice that $\hat{L}^{1/2}_1 \hat{L}^{1/2}_1  \approx \hat{L}_1$. In the first approximation, the structure constants of $e_i e_j$ vanish, and in the second, the structure constants of $e_i e_j e_k$ also vanish in the same way (with $i,j,k \neq 4$). 
Note that $\hat{G}^2$ corresponds to the operator $L_1^2$ with all first-order terms $\hat{D}_i$ removed, 
while $\hat{G}^3$ corresponds to the operator $L_1^3$ with neither the second-order operators $\hat{D}_j \hat{D}_k$ 
nor the first-order operators $\hat{D}_i$ present.

\subsection[]{Perturbed Commutations of $\hat{L}_{KG}$}

For more general quantum systems (perturbations, deformations, etc.). \begin{equation} \label{74}
    \comm{\hat{H}}{\hat{L}_n} = F(\hat{L}_n) 
\end{equation}
We want to find a first-order differential operator $\hat{L}_n$ described in terms of a basis of operators.
If we express $\hat{H}  = \hat{H}_0 + \hat{V}$
with $\hat{H}_0$ the operator of a free system and, $\hat{V}$ an interaction operator, we can use perturbation theory to solve (\ref{74}). We expand $\hat{L}_n$ as \begin{equation} \label{75}
    \hat{L}_n = \sum^M_{k=0} \epsilon^k (\hat{L}_n)_k + O(\epsilon^{M+1}),
\end{equation}
with $0<\epsilon < 1$ being a dimensionless constant.
Now, we impose \begin{equation}
    \comm{\hat{H}_0}{(\hat{L}_n)_0} = 0.
\end{equation}
Therefore, $(\hat{L}_n)_0$ is our n-th order operator, so that we can use the theory of Section \ref{sec2} with the basis and combinations of $\hat{D}_i$ that generate the centralizer $Z_n(\hat{H}_0)$. 

\textbf{First Perturbation:}
Expanding the commutator (\ref{74}) in the first perturbation $\epsilon^2 \to 0$, as
\begin{equation}\label{77}
    \comm{\hat{H}}{(\hat{L}_n)_0} + \epsilon\comm{\hat{H}}{(\hat{L}_n)_1} = F(\hat{L}_n),
\end{equation} having the left-hand side of the equation is a function of an operator, it can be expanded as an infinite sum, $F(\hat{L}_n) =\sum^\infty_{p=0} c_p ((\hat{L}_n)_0^p +p\epsilon(\hat{L}_n)^{p-1}_0(\hat{L}_n)_1) $ where $c_p \in \mathbb{C}$ and we have assumed $\comm{(\hat{L}_n)_0}{(\hat{L}_n)_1} \ket{\Phi_c} =0$, for simplicity, for some $\ket{\Phi_c} \in f(\mathbb{H}) $. From (\ref{77}) we obtain
\begin{equation} \label{78}
    \comm{\hat{V}}{(\hat{L}_n)_0} = \sum^\infty_{p=0} c_p (\hat{L}^p_n)_0,
\end{equation}

\begin{equation}\label{79}
    \comm{\hat{H}}{(\hat{L}_n)_1} = \sum^{\infty}_{p=1} c_p p (\hat{L}_n)^{p-1}_0 (\hat{L}_n)_1.
\end{equation}
Both expressions show how the interaction $\hat{V}$ acts on the unperturbed part $(\hat{L}_n)_0$ generating a power series in this operator, and how $(\hat{L}_n)_1$ evolves under $\hat{H}$ yielding another series.

\subsubsection[]{First Perturbation of $\hat{L}_{KG}$ produced by the 1-D harmonic oscillator potential }
Using a simple potential in (\ref{57}) coupled to the relativistic case of Section \ref{sec2.2.1} as the harmonic oscillator potential in one dimension $\hat{V} = kx^2$, yet assuming the first-order operator as $(\hat{L}_{KG})_1 = g_1(x) \partial_x + g_2(x)$. From (\ref{78}) and (\ref{79}), \begin{equation} 
    (\hat{L}_{KG})_1 = C_{\epsilon}e^{c_1 x} (\partial_x + c^{-1}_1 Kx) = C_{\epsilon}\hat{D}_\epsilon,
\end{equation}
where $c_{n \neq {0,1}} =0$, for consistency with the differential order in $x$; we have $C_{\epsilon} \in \mathbf{A}_{9}$. Our perturbed operator (\ref{75}) is indeed derived \begin{equation} \label{82}
    \hat{L}_{KG} = (\hat{L}_{KG})_0 + \epsilon C_\epsilon \hat{D}_\epsilon = (K^1 + \epsilon C_\epsilon e^{c_1 x})\partial_x + ... + K(1+\epsilon C_\epsilon c^{-1}_1 e^{c_1x} x).
\end{equation}
The appearance of the new $C_{\epsilon}$ breaks the super-integrability (and due to the non-commutativity assumed in \ref{74})\footnote{External perturbations or additional potentials are often responsible for this breaking, as in the perturbed harmonic oscillator or in spin models with inhomogeneous interactions. See, for example, Miller, Post, and Winternitz (2013) for a general review of classical and quantum super-integrability \cite{Miller2013}.} of (\ref{57}), since we do not assume it can be expressed as an element of the basis $\mathbf{A}_8$. 
However, it is possible to recover it in the limiting case $x \xrightarrow[]{} \infty$ if $c_1  < 0$ or $c_1 \xrightarrow{} -\infty$. Also, from (\ref{78}) we obtain the restriction equation applied to the system’s wavefunction \begin{equation}
    (2xkF^1 -c_1\hat{L}_{KG})\ket{\Phi_c} = c_0 \ket{\Phi_c}.
\end{equation}
Finally, we express the deformed Lie algebra through the commutation relations with $\hat{D}_\epsilon$. \begin{enumerate}
    \item \begin{equation}
        \comm{\hat{K}_1}{\hat{D}_\epsilon} = (c_1 +1)\hat{D}_\epsilon - e^{c_1 x} \hat{K}_1 
    \end{equation} 
    \item \begin{equation}
        \comm{\hat{O}_{1 m}}{\hat{D}_\epsilon} = -\comm{\hat{O}_{m 1}}{\hat{D}_\epsilon} = -c_1\alpha^m (\alpha^x_\mu x^\mu) \hat{D}_\epsilon + e^{c_1 x} (\alpha^x (\alpha^m_x) \hat{K}_m - c^{-1}_1 K \alpha^x_\mu x^\mu)
    \end{equation}
\end{enumerate}

Any other basis operator of $\hat{L}_{KG}$ commutes with $\hat{D}_\epsilon$, because to the sole dependence on the variable $x$, which directly deforms the linear momentum (see \ref{82}). Note that it satisfies a usual Lie structure, closed at the differential order. Nevertheless, it is not a Lie algebra of symmetry generators since $\hat{D}_{\epsilon}$ does not belong to the centralizer $Z_1(\hat{H})$. Although the perturbation of the potential breaks our symmetric structure, leading to the indeterminacy of the measurement between $\hat{L}$ and $\hat{H}$, we were still able to obtain $\hat{L}$ completely for the first first-order differential perturbation.  
To study the following differential orders, one could use the recurrence theory of orders from Section \ref{sec2}, but with the deformed Lie algebra, thus avoiding the need to perform the lengthy commutations (\ref{78}) and (\ref{79}) that arise from the general commutator (\ref{74}) for orders higher than 1, and mapping again $(\hat{L}_n)^\epsilon \sim (\hat{L}^n)^\epsilon$ but in the perturbed correspondence.

\subsubsection{The fourth-order Klein-Gordon equation}

With the first perturbation method of this section we could perform a symmetry study extension of the fourth-order Klein-Gordon equation $\hat{H}_{KG} \propto \Box^2$, hence, finding $(\hat{L}_{KG})_1$ as the last section. Such equations are studied in higher-order quantum-correction models (for example quantum gravity \cite{Moayedi2010}) or to explore new particle-properties, for instance the Higgs boson in curved space-time \cite{Cherman2012}. Writing the extended K-G equation as \begin{equation}\label{KG}
    \hat{H}^*_{KG} := a^{-2} + \Box + \frac{a^2}{2} \Box^2 + ... \frac{a^{2(N-1)}}{N!} \Box^N,
\end{equation}
expressed within \cite{Thibes2020}. In this case $a = 1/m_E$ with $m_E$ as the mass of the particle and $N \in \mathbb{N}_0$ . Now, we define $F(\hat{L}_{KG}) := \epsilon G(\hat{L}_{KG})$ (another function of $\hat{L}_{KG}$ that also could be expressed as a infinity summation) and derived the following relation
\begin{equation} \label{GC} 
   \sum^N_{n=0} \frac{a^{2(n-1)}}{n!} \comm{\Box^n}{ (\hat{L}^*_{KG})_1} = \sum^\infty_{m=0} l_m (\hat{L}^*_{KG})^m_0.
\end{equation}
In other to complete this recursive condition, we need to extend $N$ to infinity ($N \xrightarrow{} \infty$), in addition we note that $l_0 = 0$ and $l_{2n+1} = 0$ because of the commutation relation when $n=0$, and to match the differential order\footnote{Here we have assumed again that $(\hat{L}_{KG})_1$ is a first-order differential operator, but this condition is not always required as a fixed condition.} in both LHS and RHS of (\ref{GC}) respectively. Therefore, the general relation with $m = 2n $, is
\begin{equation} \label{72}
    \comm{\Box^n}{(\hat{L}^*_{KG})_1} = (\hat{L}^*_{KG})^{2n}_0 \quad (n \neq{0)},
\end{equation}
thus, the coefficients satisfy $\frac{a^{2(n-1)}}{n!} = l_{2n}$. Since our case of interest corresponds to the limited equation of (\ref{KG}), it suffices to retain only the first two elements of the derived relation (\ref{72}). The summation in (\ref{GC}) will be truncated when $2n > 4$, as this exceeds the scope required for this extension. 
Therefore, we encounter, \begin{equation}\label{DILA}
    (\hat{L}_{KG})_1 = {x}^\rho \partial_\rho,
\end{equation} 
as the first order perturbation operator which satisfies (\ref{74}). We note that the operator (\ref{DILA}) corresponds to the generator of dilations in Minkoswski space, reflecting the (new) scaling symmetry of the system. This constitutes an alternative algebraic derivation, contrasting with the conventional approach used in conformal field theory (CFT) \cite{Shaposhnikov2023}, to obtain the dilation operator (see another approach \cite{Bostelmann2010}).

Also, the non-perturbed operator $(\hat{L}_{KG})_{0}$ is restricted to $(\hat{L}_{KG})^2_{0} = 2\Box$. This interesting condition only appears when $n = 2$, and even in higher orders, and has enabled us to simplify the obtain the operator (\ref{DILA}). For example, in the Klein-Gordon equation ($n=1$) we would have had only the relation: $\comm{\Box}{(\hat{L}_{KG})_1} = (\hat{L}_{KG})^2_{0}$, who is a tough equation to solve taking count the long second-order symmetry operator (\ref{55}). Continuing with the restricted condition, a natural way to solve it, is taking $K^\mu = 2\gamma^\mu$ as the Gamma matrices, and taking the elements $ e_n e_m \in\{ K^\mu C^a, K^\mu K, C^\mu K^a, C^\mu C^a, C^\mu K, KK^a, KC^a, K^2 \} = \{0,0,0,0,0,0,0,0\}$. Hence, with $\mu,v \in (1,2,3,4)$ $  K_{\mu } = e_\mu \xrightarrow[]{} e_\mu e_v = \sum^8_{n=1} c^n_{\mu v} e_n $ and $4\gamma_{\mu -1} \gamma_{v-1} = 4\eta_{\mu -1, v-1}$, we put $c^5_{\mu v}= 4\eta_{\mu -1, v-1} $ with $e_5 =K $ being the unitary element $I$. All remaining structure constants $c^n_{pm}$ vanish. Finally, the complete first order operator $\hat{L}_{K G}$ of $\hat{H}_{KG} = a^{-2} + \Box + \frac{a^2}{2} \Box^2$ is expressed as the solution to the general commutation relation (\ref{74}), as

\begin{equation}
     K^\mu \partial_\mu + \frac{1}{4} C^\mu (\alpha^v)^{-1} \hat{O}_{v \mu} (x) +  K + \epsilon{x}^\rho \partial_\rho = (K^\mu + \epsilon x^\mu)\partial_\mu + \frac{1}{4} C^\mu (\alpha^v)^{-1} \hat{O}_{v \mu} (x) +  K.
\end{equation}
We encounter our normal symmetry approach as $\epsilon \xrightarrow[]{} 0$ in the limit. This system does not enjoy full conformal invariance due to the absence of special conformal symmetry. Instead, its symmetry algebra is given by the Weyl algebra $\mathfrak{w}(1,3)$ which is particularly relevant in theories exhibiting scale invariance without full conformal invariance \cite{Shaposhnikov2023}.

\section{Conclusion} 

By induction from (\ref{50}), the number of operators $a(n)$ for a differential basis of order $n$, belonging to $Z_n(\hat{H})$, with $ dim \mathbf{A}_N = N$ (including the unit operator), that arise after the commutation with the central equation $\hat{H}$ and $\hat{L}_n$, is \begin{equation}
    a_N(n) = \frac{(n+N-1)!}{(N-1)! n!}.
\end{equation}
When $n=1$, this is the number of basis operators of the system’s symmetry generators \begin{equation}
    a_N(1) = \frac{(N)!}{(N-1)!} = N.
\end{equation}
Let $e_\mu \in \mathbf{A}_N = \mathbf{A}^+_N \oplus\mathbf{A}^-_N$. We have constructed the projection function to the basis of symmetry generators \begin{equation}
    {e_{i_1} e_{i_2} ... e_{i_n}}\in G_{(N)}(\mathbf{A}_{a_N(n)}) = G^n(\mathbf{A}_N) \subset \mathbf{A}_N;
\end{equation}
from which \begin{equation}
G^n(\mathbf{A}_N) = G(\underbrace{\mathbf{A}_N, \mathbf{A}_N, \dots, \mathbf{A}_N}_{n \text{ times}})
\end{equation}
We have shown that it is possible to systematically obtain the centralizers' basis of different orders and to establish their recursive relation through the proposed algebraic structure. Here are two key generalizations: First, defining the basis \(\mathbf{A}_N\) with built-in cyclic operations ensures the linear independence of higher-order operators, which all modifications to operations and their higher sequences stem solely from selecting complex structure constants, without this, removing an accompanying element from a base operator \(\hat{D}\) leads to combinations that impose unwanted constraints on the constants of other \(\hat{L}^n\) operators, creating undesirable dependencies. Our formulation allows any basis element (and its associated symmetry generator) to be eliminated simply by choosing the appropriate structure constants. Second, the Lie reduction, enables an exact mapping between the operators produced by the sequence \(\hat{L}^n\) and the higher-order operators obtained after commuting \(\hat{L}_n\) with \(\hat{H}\). Without this mapping, there is no precise correspondence between \(\hat{L}^n\) and \(\hat{L}_n\). Thankfully, the Lie algebra’s central-operator structure arranges them exactly as required. Together, these two generalizations complete the independent construction of higher-order operators.

We also saw that, when the commutation with the central operator is not null, perturbation theory allows the construction of operators that respect modified relations. This approach provides a compact tool to study symmetries and centralizers in quantum mechanics, both in ideal cases and in possible perturbed scenarios.

\section{Acknowledgments} This article was inspired by the work of Prof. Nikolay Sukhomlin.

\newpage
\appendix

\section[]{Algebraic Basis of $\hat{L}_1$ and $\hat{L}_{KG}$} \label{ap1}

Let $\mathbf{A}_N \subset \mathcal{D}$ be an algebra with a bilinear product $xy$. The product is separable $xy = xy_+ + xy_-$, with Jordan product $\displaystyle xy_+ = \{x ,y\}$ and Lie product $\displaystyle xy_- = \comm{x}{y}$ . The Jordan product is the analogue of the anti-commutator, and is therefore symmetric, $xy_+ = yx_+$. The Lie product is the analogue of the commutator and is anti-symmetric, $xy_- = -yx_-$. We will denote these sub-algebras generated by these two products as $\mathbf{A_N^+}$ and $\mathbf{A_N^-}$ respectively. Note that $N=$ Dim $\mathbf{A}_N = $ Dim $\mathbf{A_N^+} =$ Dim $ \mathbf{A_N^-}$. Finally, this algebra $\mathbf{A}_N$ is taken to be associative. 
Now, expressing the multiplication in terms of the elements of an algebraic basis in $\mathbf{A}_N$,  i.e: $e_i e_j = \sum^N_{k=1} c^k_{i j} e_k$ with $ c^{k}_{i j} \in \mathbb{C}$, we decompose into
\begin{itemize}
    
    \item[] Symmetric product \begin{equation} \label{eq17}
        \comm{e_i}{e_j}_+ = \sum^N_{k=1} (c^k_{i j} + c^k_{j i})e_k \doteq \sum^N_{k=1}(c^k_{i,j})^+e_k
    \end{equation}
    \item[] Anti-symmetric product \begin{equation} \label{eq18}
        \comm{e_i}{e_j} = \sum^N_{k=1} (c^k_{i j} - c^k_{j i})e_k \doteq \sum^N_{k=1}(c^k_{i,j})^- e_k.
    \end{equation}
\end{itemize}
Finally, our construction ensures a symmetric and anti-symmetric algebra over the set of functions $F(x,t)_i \in \Xi(F) $ expressed in the above basis in the Lie algebra and the Jordan algebra, both associative and flexible algebras \cite{Okubo} \cite{McCrimmon2004}. Since any associative algebra is an alternative algebra, and since the composition algebras, quaternions, and octonions are both quadratic alternative algebras, we can construct an isomorphic mapping to any algebra of this type. It is easy to see that we can move to the Clifford algebra $\mathbf{C}^n$ if the symmetric basis satisfies $\{e_i ,{e_j}\} = e_i e_j + e_j e_i = -2\delta_{i j} \mathbf{I} $. The (N-1)-imensional Clifford algebra is known to have a close relation with the group $SO(N-1)$. For example, the octonions are intimately related to the group $SO(7)$ \cite{Baez2002}. 

Now let $\hat{A}, \hat{B}\in Z_1(\hat{H})$, We can express the multiplication of first-order operators, taking into account that if $ord(L_1) = m$ it has constant elements $e_p$, en función de \textit{Jordan-álgebra-valued differential operators} in terms of \textit{Jordan-algebra-valued differential operators} and \textit{Lie-algebra-valued differential operators} of first order as
\begin{equation} \label{eq20}
    \hat{A} \hat{B} = \sum^{ord(\hat{L}_{1})}_{i=1} \sum^{ord(\hat{L}_1)}_{j=1} a_i b_j\hat{D}_{i}\hat{D}_{j}  \in Z_2(H).
\end{equation}

It is then separable.
\begin{equation} \label{eq21}
    \hat{A}\hat{B} = \hat{A}\hat{B}_+ + \hat{A}\hat{B}_- =  \frac{1}{2}\sum^{ord(\hat{L}_1)}_{i=1} \sum^{ord(\hat{L}_1)}_{j=1} \{a_i ,{b_j}\}\hat{D}_{i} \hat{D}_{j} + \frac{1}{2}\sum^{ord(\hat{L}_1)}_{i=1} \sum^{ord(\hat{L}_1)}_{j=1} \comm{a_i}{b_j} \hat{D}_{i} \hat{D}_{j} 
\end{equation}

\subsection[]{Inverse sub-algebra $\mathbf{A}_{N,-1}$} \label{inv}
We introduce the notion of an inverse element within the algebraic sub-algebra $\mathbf{A}_{N,-1} \subset \mathbf{A}_N$. This construction provides the 
appropriate framework to characterize how inversion operates in our setting. 
First, we assume that there exists an element $k_i$ such that $k_i e_i = e_N$, 
where we denote by $e_N$ the central unit operator satisfying $\comm{e_N}{e_i}=0$. 
If $k_i$ can be expressed as a linear combination of constants, 
$k_i = \sum^N_1 a^{(j)}_{i} e_j \in \mathbf{A}_N$, 
with $a^{(j)}_{i} \in \mathbb{C}$, 
then we can denote $k_i = e^{-1}_i$ with the operation
\begin{equation}
    e^{-1}_i e_i = e_i e^{-1}_i = \sum_j \sum_n a^{(j)}_{i} c^n_{j j} e_n.
\end{equation}
To satisfy this condition, we must require that 
$\sum_j \sum_n a^{(j)}_{i} c^n_{j j} = 0$ when $n \neq N$, 
and $\sum_j a^{(j)}_{i} c^N_{j j} = 1$.

\newpage

\section[]{Detailed Expressions of the Third-Order Basis Differential Operators of $\hat{L}_3$}\label{A}
This appendix provides the explicit forms of the third-order partial differential operators used in Section \ref{sec2.1.5} as the basis of $\hat{L}_3$. 

\begin{table}[ht]
\centering
\begin{tabular}{|c|}
\hline
$\hat{D}_{3,1} = -2i \frac{t^3}{3} \partial^3_x - (t^2x + i\frac{t^3}{3})\partial^2_x + \frac{i}{2}(tx+ it^2x - \frac{ t^3}{3} +it^2) \partial_x + \frac{i}{4}(\frac{-ix^3}{3}+ tx^2 + it^2x -\frac{ t^3}{3})$ \\
\hline
$\hat{D}_{3,2} = -2it^2 \partial_x^3 - (2tx + it^2)\partial_x^2 + \frac{i}{2}(x^2 + 2itx -t^2 +2it) \partial_x + \frac{i}{4} (x+it)^2$ \\

\hline
$\hat{D}_{3,3} = -4i t\partial^3_x - x\partial^2_x + i\partial_x$ \\
\hline
$\hat{D}_{3,4} = -4i\partial^3_x$ \\
\hline
$\hat{D}_{3,5} = 2it\partial^3_{x,t,t} +x\partial^2_x$ \\
\hline
$\hat{D}_{3,6} = it^2 \partial^3_{x,x,t} + tx\partial^2_{x,t} - \frac{i}{4}(x^2 + 2it) \partial_t$ \\
\hline
$\hat{D}_{3,7} = 2it\partial^3_{x,x,t} + x\partial^2_{x,t}$ \\
\hline
$\hat{D}_{3,8} = -i\partial^3_t$ \\
\hline
$\hat{D}_{3,9} = -i\partial^3_{x,x,t}$ \\
\hline
$\hat{D}_{3,10} = i\partial^3_{x, t, t}$ \\
\hline
$\hat{D}_{3,11} = -2it\partial^2_x - (x+it)\partial_x - \frac{1}{2}(x+it)$ \\
\hline
$\hat{D}_{3,12} = -2t\partial^2_{x} + i\partial_x + \frac{i}{2}$ \\
\hline
$\hat{D}_{3,13} = t\partial^2_{x, t} - \frac{i}{2}x\partial_t$ \\
\hline
$\hat{D}_{3,14} = \partial^2_{x}$ \\
\hline
$\hat{D}_{3,15} = \partial^2_t$ \\
\hline
$\hat{D}_{3,16} = \partial^2_{x,t}$ \\
\hline
$\hat{D}_{3,17} = 2it\partial_{x} - x$ \\
\hline
$\hat{D}_{3,18} = i\partial_t$ \\
\hline
$\hat{D}_{3,19} = -i\partial_{x}$ \\
\hline
$\hat{D}_{3,20} = \hat{I}$ \\
\hline
\end{tabular}
\caption{Third order differential operators resulting from the commutation with $\hat{H}$ and $\hat{L}_3$}
\label{apptab:operators}
\end{table}

\section[]{$\hat{L}^4_1$ in Compact Basis Form}\label{apC}
\begin{equation}\label{eq:L1L3_to_L4}
\begin{aligned}
\hat{L}_1^{4}
&= (e_4 e_4 e_4)\,\hat{L}_1^{3}
+ \sum_{v=1}^{4} \Bigg\{
\sum_{u=1}^{4}\sum_{i=1}^{3}
\Big[
(e_i e_i e_u e_v)\,\hat{D}_i^2 \hat{D}_u \hat{D}_v
+ \sum_{n=1}^{4} (e_i e_i e_u e_v)\,C^{n}_{u v}\,\hat{D}_i^2 \hat{D}_{n} \\
&\qquad\qquad\qquad\quad
+ \sum_{k=1}^{4} (e_i e_i e_u e_v)\,C^{k}_{i v}\,\hat{D}_i \hat{D}_{k} \hat{D}_{u}
+ \sum_{k=1}^{4} (e_i e_i e_u e_v)\,C^{k}_{i v}\,\hat{D}_{k} \hat{D}_{i} \hat{D}_{u} \\
&\qquad\qquad\qquad\quad
+ \sum_{k=1}^{4}\sum_{n=1}^{4} (e_i e_i e_u e_v)\,C^{k}_{i v} C^{n}_{i k}\,\hat{D}_{n}\hat{D}_{u}
+ \sum_{k=1}^{4}\sum_{n=1}^{4} (e_i e_i e_u e_v)\,C^{n}_{u v} C^{k}_{i n}\,\hat{D}_{i} \hat{D}_{k}
\Big] \\ &\quad +
\sum_{u=1}^{4}\sum_{i<j}^{4}
\Big[
\{e_i,e_j\} e_u e_v\,\hat{D}_i \hat{D}_j \hat{D}_u \hat{D}_v
+ \sum_{n=1}^{4} \{e_i,e_j\} e_u e_v\,C^{n}_{u v}\,\hat{D}_i \hat{D}_j \hat{D}_{n} \\
&\qquad\qquad\qquad\quad
+ \sum_{k=1}^{4} \{e_i,e_j\} e_u e_v\,C^{k}_{i v}\,\hat{D}_j \hat{D}_u \hat{D}_{k}
+ \sum_{k=1}^{4} \{e_i,e_j\} e_u e_v\,C^{k}_{j v}\,\hat{D}_i \hat{D}_u \hat{D}_{k} \\
&\qquad\qquad\qquad\quad
+ \sum_{k=1}^{4}\sum_{n=1}^{4} \{e_i,e_j\} e_u e_v\,C^{k}_{i v} C^{n}_{j k}\,\hat{D}_{n} \hat{D}_{u}
+ \sum_{k=1}^{4}\sum_{n=1}^{4} \{e_i,e_j\} e_u e_v\,C^{k}_{j v} C^{n}_{i k}\,\hat{D}_{n} \hat{D}_{u} \\
&\qquad\qquad\qquad\quad
+ \sum_{k=1}^{4}\sum_{n=1}^{4} \{e_i,e_j\} e_u e_v\,C^{n}_{u v} C^{k}_{i n}\,\hat{D}_{j} \hat{D}_{k}
+ \sum_{k=1}^{4}\sum_{n=1}^{4} \{e_i,e_j\} e_u e_v\,C^{n}_{u v} C^{k}_{j n}\,\hat{D}_{i} \hat{D}_{k}
\Big] \\ &\quad +
\sum_{u=1}^{4}\sum_{i<j}^{4}\sum_{k=1}^{4}
\Big[
C^{k}_{ij}(e_i e_j e_u e_v)\,\hat{D}_u \hat{D}_k \hat{D}_v
+ \sum_{n=1}^{4} C^{k}_{ij}(e_i e_j e_u e_v)\,C^{n}_{k v}\,\hat{D}_u \hat{D}_{n} \\
&\qquad\qquad\qquad\quad
+ \sum_{n=1}^{4} C^{k}_{ij}(e_i e_j e_u e_v)\,C^{n}_{u v}\,\hat{D}_{n} \hat{D}_{k}
\Big] \\ &\quad +
\sum_{u=1}^{4}\sum_{i<j}^{4}\sum_{k=1}^{4}\sum_{n=1}^{4}
\Big[
C^{k}_{ji} C^{n}_{k u}(e_i e_j e_u e_v)\,\hat{D}_{n} \hat{D}_{v}
+ \sum_{p=1}^{4} C^{k}_{ji} C^{n}_{k u}(e_i e_j e_u e_v)\,C^{p}_{n v}\,\hat{D}_{p}
\Big]
\Bigg\}.
\end{aligned}
\end{equation}


\begin{thebibliography}{99}

\bibitem{Bokhari2010}
Bokhari, A. H., Mahomed, F. M., \& Zaman, F. D. (2010).
Symmetries and integrability of a fourth-order Euler–Bernoulli beam equation.
\textit{Journal of Mathematical Physics}, 51, 053517.
\href{https://doi.org/10.1063/1.3377045}{https://doi.org/10.1063/1.3377045}



\bibitem{Busch2013}
Busch, P., Lahti, P., \& Werner, R. F.
\newblock Proof of Heisenberg’s Error–Disturbance Relation.
\newblock \textit{Physical Review Letters}, \textbf{111}, 160405, 2013.
\newblock \href{https://doi.org/10.1103/PhysRevLett.111.160405}{doi:10.1103/PhysRevLett.111.160405}.

\bibitem{Miller2013}
Miller, W., Post, S., \& Winternitz, P.
\newblock Classical and Quantum Superintegrability with Applications.
\newblock \textit{Journal of Physics A: Mathematical and Theoretical}, \textbf{46}(42), 423001, 2013.
\newblock \href{https://doi.org/10.1088/1751-8113/46/42/423001}{doi:10.1088/1751-8113/46/42/423001}.

\bibitem{Soper2011}
Soper, D. E.
\newblock Galilean Boost Symmetry.
\newblock University of Oregon, April 2011.
\newblock \url{https://pages.uoregon.edu/soper/QuantumMechanics/boosts.pdf}.

\bibitem{Hestenes1966}
Hestenes, D.
\newblock \textit{Space--Time Algebra}.
\newblock 2nd ed., Springer, 2015.
\newblock doi: \href{https://doi.org/10.1007/978-3-319-18413-5}{\nolinkurl{10.1007/978-3-319-18413-5}}.

\bibitem{Nikitin2016}
A.~G.~Nikitin,
``Higher-order symmetries for linear and nonlinear Schrödinger equations,''
doi:\href{https://arxiv.org/abs/1603.01715}{arXiv:1603.01715}.





\bibitem{Franchino2022}
Franchino-Viñas, S.~A., \& Relancio, J.~J.
\newblock Geometrizing the Klein--Gordon and Dirac equations in Doubly Special Relativity.
\newblock \textit{Classical and Quantum Gravity}, \textbf{39}(21), 215004, 2022.
\newblock doi: \href{https://doi.org/10.1088/1361-6382/ac8f3b5}{\nolinkurl{10.1088/1361-6382/ac8f3b5}}.

\bibitem{Zheglov2011}
Zheglov, A.~B.
``On rings of commuting partial differential operators''.
\textit{Sbornik: Mathematics}, \textbf{202}(7), 1031--1061, 2011.
\href{https://www.ams.org/smj/2014-25-05/S1061-0022-2014-01316-7/}{doi:10.1070/SM2011v202n07ABEH004179}.




\bibitem{Sakai1971}
Sakai, S.
\newblock \emph{C$^{\ast}$-Algebras and W$^{\ast}$-Algebras}.
\newblock Springer, Berlin, 1971.
\href{https://link.springer.com/book/10.1007/978-3-642-61993-9}{doi:10.1007/978-3-642-61993-9}.



\bibitem{Halvorson2006}
Halvorson, H., \& Müger, M. (2006).
\textit{Algebraic Quantum Field Theory}.
En J. Butterfield \& J. Earman (Eds.), \textit{Handbook of the Philosophy of Physics} (pp. 731--922).
Elsevier.
doi:\href{https://doi.org/10.1016/B978-044451560-5/50010-4}{10.1016/B978-044451560-5/50010-4}



\bibitem{Iqbal2018}
Iqbal, M., \& Zhang, Y. (2018).
Lie Symmetries of Klein-Gordon and Schrödinger Equations.
\textit{Applied Mathematics}, 9, 336–346.
\href{https://www.scirp.org/journal/paperinformation.aspx?paperid=84566}{https://doi.org/10.4236/am.2018.94025}


\bibitem{Jordan1934}
Jordan, P., von Neumann, J., \& Wigner, E. (1934).
On an algebraic generalization of the quantum mechanical formalism.
\textit{Annals of Mathematics}, 35(1), 29–64.

\bibitem{Kasman2001}
Kasman, A., \& Previato, E. (2001).
Commutative partial differential operators.
\textit{Physica D: Nonlinear Phenomena}, 152–153, 66–77.
\href{https://www.sciencedirect.com/science/article/pii/S0167278901001724}{https://doi.org/10.1016/S0167-2789(01)00172-4}

\bibitem{Pain2013}
J.-C.~Pain,
\newblock ``Commutation relations of operator monomials'',
\newblock {\em Journal of Physics A: Mathematical and Theoretical}, \textbf{46}(3), 035304, 2013.
\newblock doi:\href{https://doi.org/10.1088/1751-8113/46/3/035304}{10.1088/1751-8113/46/3/035304}.


\bibitem{Baez2002}
Baez, J.~C.
\newblock The octonions.
\newblock \emph{Bulletin of the American Mathematical Society}, 39(2):145--205, 2002.
\newblock doi: \href{https://doi.org/10.1090/S0273-0979-01-00934-X}{\nolinkurl{10.1090/S0273-0979-01-00934-X}}.

\bibitem{Okubo}
Okubo, S.
\textit{Introduction to Octonion and Other Non-Associative Algebras in Physics}.
Department of Physics and Astronomy, University of Rochester, NY. 

\bibitem{McCrimmon2004}
McCrimmon, K.
\newblock \textit{A Taste of Jordan Algebras}.
\newblock Universitext. Springer-Verlag, New York, 2004.
\newblock doi: \href{https://doi.org/10.1007/b97489}{\nolinkurl{10.1007/b97489}}.
\newblock ISBN: 978-0-387-95447-9.

\bibitem{JimenezPastor2025}
Jiménez-Pastor, A., Rueda, S.~L., Zurro, M.~A., Hernández-Heredero, R., \& Delgado, R. (2025).
Computing almost commuting bases of ODOs and Gelfand--Dickey hierarchies.
\textit{Mathematics in Computer Science}, 19, 4.
\href{https://link.springer.com/article/10.1007/s11786-025-00601-9}{doi:10.1007/s11786-025-00601-9}.



\bibitem{Putnam}
Putnam, I.~F.
\textit{C$^*$-algebras and Topological Dynamics: From Dynamical Systems to Operator Algebras}.
University of Victoria.
\href{https://web.uvic.ca/~ifputnam/r/Wayne_colloquium_1.pdf}{Disponible en línea}.



\bibitem{Sukhomlin2004}
Sukhomlin, N., \& Arias, M. (2004).
Estudio de Simetría y de Posibilidades de la resolución exacta de las ecuaciones de Schrödinger y Hamilton–Jacobi para un Sistema Aislado, Primera Parte: Clasificación de los operadores de Simetría hacia Tercer Orden.
\textit{Ciencia y Sociedad}, XXIX(1), 1–23.

\bibitem{Zhang2009}
Zhang, F., \& Chen, J. (2009).
Dynamical symmetries of the Klein-Gordon equation.
\textit{Journal of Mathematical Physics}, 50(3), 032901.
\href{https://pubs.aip.org/aip/jmp/article/50/3/032901/391928}{https://doi.org/10.1063/1.3081615}


\bibitem{Folland1989}
Folland, G.~B.
\textit{Harmonic Analysis in Phase Space}.
Princeton University Press, Princeton, NJ, 1989.
Annals of Mathematics Studies, Vol.~122.
doi: \href{https://doi.org/10.1515/9781400882427}{\nolinkurl{10.1515/9781400882427}}.
ISBN: 978-0-691-08527-5.


\bibitem{deGosson2006}
de~Gosson, M.~A.
\textit{Symplectic Geometry and Quantum Mechanics}.
Birkh{\"a}user, Basel, 2006.
Operator Theory: Advances and Applications, Vol.~166.
doi: \href{https://doi.org/10.1007/3-7643-7575-2}{\nolinkurl{10.1007/3-7643-7575-2}}.
ISBN: 978-3-7643-7574-4.


\bibitem{Hall2013}
Hall, B.~C. 
\textit{Quantum Theory for Mathematicians}.
Springer, New York, 2013.
Graduate Texts in Mathematics, Vol.~267.
doi: \href{https://doi.org/10.1007/978-1-4614-7116-5}{10.1007/978-1-4614-7116-5}.
ISBN: 978-1-4614-7115-8.

\bibitem{weinbergQTF1}
S.~Weinberg,
\newblock \emph{The Quantum Theory of Fields, Vol.~I: Foundations}.
\newblock Cambridge University Press, 1995.
\newblock doi:\href{https://doi.org/10.1017/CBO9781139644167}{10.1017/CBO9781139644167}

\bibitem{greinerFQ}
W.~Greiner,
\newblock \emph{Field Quantization}.
\newblock Springer, 1996.
\newblock doi:\href{https://doi.org/10.1007/978-3-642-61480-1}{10.1007/978-3-642-61480-1}

\bibitem{BreuerPetruccione2002}
H.-P.~Breuer y F.~Petruccione,
\newblock \emph{The Theory of Open Quantum Systems}.
\newblock Oxford University Press, 2002.
\newblock doi:\href{https://doi.org/10.1093/acprof:oso/9780199213900.001.0001}{10.1093/acprof:oso/9780199213900.001.0001}

\bibitem{TranstrumVanHuele2005}
M.~K.~Transtrum and J.-F.~S.~Van~Huele,
``Commutation relations for functions of operators'',
\textit{Journal of Mathematical Physics}, \textbf{46}(6), 063510, 2005.
url:\href{https://pubs.aip.org/aip/jmp/article/46/6/063510/2259460}{https://pubs.aip.org/aip/jmp/article/46/6/063510/2259460}.

\bibitem{AmelinoCamelia2002}
G.~Amelino-Camelia,
``Relativity in space-times with short-distance structure governed by an observer-independent (Planckian) length scale,''
\emph{International Journal of Modern Physics D}, vol.~11, no.~1, pp.~35--60, 2002.  
doi: \href{https://doi.org/10.1142/S0218271802001330}{10.1142/S0218271802001330}

\bibitem{Chaichian2004}
M.~Chaichian, P.~Kulish, K.~Nishijima, and A.~Tureanu,
``On a Lorentz-invariant interpretation of noncommutative space-time and its implications on noncommutative QFT,''
\emph{Physics Letters B}, vol.~604, pp.~98--102, 2004.  
doi: \href{https://doi.org/10.1016/j.physletb.2004.10.045}{10.1016/j.physletb.2004.10.045}

\bibitem{Majid1994}
S.~Majid and H.~Ruegg,
``Bicrossproduct structure of $\kappa$-Poincaré group and noncommutative geometry,''
\emph{Physics Letters B}, vol.~334, pp.~348--354, 1994.  
doi: \href{https://doi.org/10.1016/0370-2693(94)90699-8}{10.1016/0370-2693(94)90699-8}



\bibitem{Stampfli1961}
J.~G.~Stampfli,
\newblock ``Commutators, perturbations, and unitary spectra'',
\newblock {\em Acta Mathematica}, \textbf{106}, 297--306, 1961.
\newblock doi:\href{https://doi.org/10.1007/BF02545787}{10.1007/BF02545787}.

\bibitem{Hirata2018}
C. M. Hirata,
``Lecture Notes on General Relativity: Symmetries and Killing Vectors,''
Ohio State University (2018).
Available at: \url{https://cosmo.nyu.edu/yacine/teaching/GR_2018/lectures/lecture19.pdf}

\bibitem{Fu2023}
Y.~Fu, X.~Gu, Y.~Wang, W.~Cai,
``Mass-, and Energy Preserving Schemes with Arbitrarily High Order for the Klein--Gordon--Schr\"odinger Equations,''
\textit{Journal of Scientific Computing} \textbf{97}, 75 (2023).
\href{https://doi.org/10.1007/s10915-023-02388-y}{doi:10.1007/s10915-023-02388-y}

\bibitem{Cherman2012}
A.~Cherman, L.G.~Ferreira Filho, L.L.~Santos Guedes, J.A.~Helayël-Neto,
“Explicit classical solutions and comments on Higher-Derivative Klein-Gordon equation in (1+1)-D,”
\emph{Rev. Mex. Fís.} **58**(5) (2012) 384–390.
\href{https://www.redalyc.org/pdf/570/57025089009.pdf}{https://www.redalyc.org/pdf/570/57025089009.pdf}

\bibitem{Thibes2020}
R.~Thibes,
``Natural Higher-Derivatives Generalization for the Klein–Gordon Equation,''
arXiv:2011.02567 [math-ph] (2020).  
\href{https://arxiv.org/abs/2011.02567}{arXiv:2011.02567 [math-ph]}

\bibitem{Moayedi2010}
S.K.~Moayedi, M.R.~Setare, H.~Moayeri,
“Quantum Gravitational Corrections to the Real Klein-Gordon Field in the Presence of a Minimal Length,”
\emph{Int. J. Theor. Phys.} **49**(9) (2010) 2080–2088.
\href{https://link.springer.com/article/10.1007/s10773-010-0394-2}{https://link.springer.com/article/10.1007/s10773-010-0394-2}

\bibitem{Bostelmann2010}
H.~Bostelmann, C.~D’Antoni, G.~Morsella,
“On Dilation Symmetries Arising from Scaling Limits,”
\emph{Communications in Mathematical Physics}, Vol.\ 294 (2010), pp.\ 21-60.
\href{https://link.springer.com/article/10.1007/s00220-009-0899-9}{https://link.springer.com/article/10.1007/s00220-009-0899-9}

\bibitem{Shaposhnikov2023}
M.~Shaposhnikov, A.~Tokareva,
“Exact quantum conformal symmetry, its spontaneous breakdown, and gravitational Weyl anomaly,”
\emph{Physical Review D}, Vol.107, No.6 (2023), p.065015.
\href{https://doi.org/10.1103/PhysRevD.107.065015}{https://doi.org/10.1103/PhysRevD.107.065015}


\bibitem{Frank2007}
R.~L. Frank, E.~H. Lieb and R. Seiringer, ``Hardy–Lieb–Thirring inequalities for fractional Schrödinger operators'', *Journal of the American Mathematical Society*, vol. 20, pp. 1075-1100, 2007. \href{https://arxiv.org/abs/math/0610593}{https://arxiv.org/abs/math/0610593}

\bibitem{Valverde2014}
L.~A. Valverde, ``Trace asymptotics for fractional Schrödinger operators'', *Journal of Mathematical Analysis and Applications*, vol. 417, pp. 171-191, 2014. \href{https://www.sciencedirect.com/science/article/pii/S0022123613004199}{https://www.sciencedirect.com/science/article/pii/S0022123613004199}

\bibitem{Obukhov2021}
V. V. Obukhov, ``Algebra of symmetry operators for Klein–Gordon-Fock equation'', *Symmetry*, vol. 13, no. 4:727, 2021. \href{https://doi.org/10.3390/sym13040727}{https://doi.org/10.3390/sym13040727}

\bibitem{Miraboutalebi2022}
S. Miraboutalebi, ``Effect of RGUP on the nonlinear Klein-Gordon model with self-interaction'', *Physics Letters B*, vol. 826, 2022. \href{https://www.sciencedirect.com/science/article/pii/S037026932200404X}{https://www.sciencedirect.com/science/article/pii/S037026932200404X}






\end{thebibliography}
\end{document}